\documentclass[aps,preprintnumbers,amsmath,amssymb,nofootinbib,eqsecnum, preprintnumbers]{revtex4}
\usepackage{eurosym}
\usepackage{amsfonts}
\usepackage{amsmath}
\usepackage{amssymb,epsf}
\usepackage{color}
\usepackage{graphicx}
\usepackage{natbib}
\usepackage{float}
\usepackage{caption}
\usepackage{subfig}
\usepackage{epstopdf}

\begin{document}
\title{A Study of Black Holes in $F(R)-$ModMax Gravity: Gravitational Lensing and Constraints from EHT Observations}
\author{Khadije Jafarzade$^{1}$\footnote{
email address: khadije.jafarzade@gmail.com}, Zeynab Bazyar$^{2}$\footnote{
email address: hzp.bazyar@gmail.com}, Mubasher Jamil$^{3}$\footnote{
email address: mjamil@sns.nust.edu.pk}}
\affiliation{$^{1}$Department of Theoretical Physics, Faculty of Science, University of Mazandaran,
P. O. Box 47416-95447, Babolsar, Iran\\
$^{2}$Department of Physics, Isfahan University of Technology, 84156-83111,
Isfahan, Iran\\
$^{3}$School of Natural Sciences, National University of Sciences and Technology, 44000 Islamabad, Pakistan}
\begin{abstract}
The study of astrophysical phenomena like black hole shadows is an effective approach to properly understand the modified gravity and explore its validity. Motivated by recent astrophysical observations, we consider a black hole (BH) in $F(R)-$ModMax gravity and study the optical features such as the shadow's geometrical shape,  energy emission rate, and deflection of light. More specifically, we show how the variation of the model parameters imprints specific signatures on these optical quantities. In the following, we consider such black holes as supermassive BHs and evaluate the parameters of the model with shadow size estimates done by the observations of M87*  from the Event Horizon Telescope (EHT). According to our findings, the parameter $f_{R_{0}}$ plays an effective role in having results consistent with the EHT data such that the resulting shadow of AdS black holes in $F(R)-$ModMax gravity agrees with the observational data for $f_{R_{0}}<-1$. However, for dS black holes, a consistent result is observed for $f_{R_{0}}>-1$.

\end{abstract}

\maketitle

\section{Introduction}
Einstein’s general relativity (GR) has been the most successful and a widely accepted theory to explain gravitational phenomena such as the precession of the perihelion advance of Mercury, the gravitational Doppler effect and deflection of light near massive stars and black holes (BHs) \cite{Will:1a}. GR has also predicted the existence of BHs and gravitational waves which have been recently experimentally verified by the LIGO-Virgo collaboration \cite{Abbott:1a,Abbott:1b,Abbott:1c}.  Despite these successes, GR fails to address recent observations such as the accelerated expansion of the universe and the dynamics of cosmic structures \cite{Riess:1a,Perlmutter:1a}. To explain both cases, we need to introduce dark energy, which is an unusual form of matter energy characterized by a substantial negative pressure. In this regard, various modified gravity models have been proposed to interpret the dark sector of the Universe such as Lovelock gravity \cite{Lovelock:1a,Lovelock:1b}, massive gravity \cite{deRham:1a}, scalar-tensor theories \cite{Brans:1a,Barrabes:1a,Eiroa:1a}, and $F(R)$ gravity theories \cite{Akbar:1a,Souza:1a,Jafarzade:2a}. The theory of $F(R)$ gravity is a straightforward modification of GR, where the scalar curvature $R$ is substituted for an arbitrary function of $R$ \cite{Akbar:1b,Cognola:1a}. Some motivations to consider $F(R)$ gravity include (i) this theory can explain the accelerated expansion and structure formation of the Universe without considering dark energy or dark matter \cite{Perlmutter:1a,Riess:1a}, (ii) such a theory coincident with Newtonian and post-Newtonian approximations \cite{Capozziello:1a,Capozziello:1b}, (iii) the action of $F(R)$ theory can encapsulate some of the basic properties of higher-order gravity \cite{Hendi:1a}, (iv) the fatal Ostrogradski instability can be avoided in this theory \cite{Woodard:1a}.

Nonlinear electrodynamics (NED) as a generalization of Maxwell's theory was first proposed by Born and Infeld in the 1930s. They formulated a theory that can remove the central singularity of the electromagnetic field of a point charge and its energy divergence \cite{Born:1a}.
In recent years, a new nonlinear model called Modified Maxwell electrodynamics (ModMax) was discovered that reduces to Maxwell's theory in the non-interacting limit, namely when the strength of self-interaction goes to zero \cite{Bandos:1a}. ModMax was proved to be the only theory that preserves both symmetries of Maxwell’s original theory: conformal invariance and electromagnetic duality. The ModMax theory is of great interest as such symmetries can lead to novel observable implications when the theory is quantized \cite{Cian:1a}. Although this theory has been studied extensively at the classical level with applications in strongly coupled condensed matter systems, it remains largely untouched in a quantum context. This novel theory has attracted extensive attention recently, ranging from applications of black hole physics in investigating the thermodynamic structure of charged black holes \cite{Amirabi:1a}, magnetic black holes \cite{Kruglov:1a}, shadow and quasinormal modes \cite{Pantig:1a},   4D Lovelock theory coupled to ModMax \cite{Amirabi:1b}, $F(R)$ gravity coupled to ModMax \cite{Panah:1b},  accelerated black holes in ModMax \cite{Barrientos:1a}. 

A century after the discovery of the first gravitational lenses, the Event Horizon Telescope (EHT) Collaboration reported the first-ever event-horizon-scale images of a black hole as predicted by general relativity \cite{EHT:1a,EHT:1b}. These images are based on the intuition that if light rays pass near a BH, the rays will bend even move in a circular orbit close to it, known as strong gravitational lensing \cite{Gralla:1a,nagina,shafia}. The graphs show the angular resolution of the supermassive black holes, seen as a dark disk in the sky. This disk contains a circular dark interior, known as the shadow, and an adjacent bright ring interpreted as a photon sphere \cite{Gan:1a}. In fact, the strong gravitational deflection of light is the origin of the shadow and photon sphere. Hence it is expected that the existing images of the black hole's shadow provide us with important information about the spacetime geometry around the black hole in the vicinity of the event horizon \cite{Bozza:1a}. According to the reported results, the BH shadow not only gives us information concerning jets and matter dynamics around black holes but can also be considered a useful tool for comparing alternative theories of gravity with general relativity. The BH shadow and gravitational lensing  have become a hot topic among researchers and has been extensively studied in black hole physics such as regular BHs \cite{Kumaran:1a,Gomez:40}, holonomy corrected BHs \cite{Soares:108,Soares:4004}, BHs in modified GR \cite{Amarilla:1a,Kumar:1ab,Jafarzade:1a,Zhu:1a, Guo:588,Asukula:109,Zeng:2020dco,Zeng:2023dco}, BHs in Perfect Fluid Dark Matter \cite{Haroon:1a}, BHs in an expanding universe \cite{Guo:101}, BHs in the presence of nonlinear electrodynamics \cite{Allahyari:1a,Jafarzade:1b}, and BHs surrounded by plasma \cite{Ahmedov:1b,Pahlavon:1a,Atamurotov:1a,Kala:84,Kala:137}, gravitational lensing by naked singularities \cite{Virbhadra:65,Virbhadra:77,Virbhadra:106}. Recently, horizon-scale emission from the black hole in the galaxy M87* and Sgr A* has been investigated using ground-based interferometry \cite{Gralla:1a}. A precise measurement of the photon ring shape is employed to infer black hole parameters and a test of GR. After observing images of supermassive black holes in galaxy M87* and Sgr A*, many people became interested in calculating the shadows of BH solutions and the confrontation with the extracted information from the EHT BH shadow image of M87* and Sgr A* \cite{Davoudiasl:1c,Jafarzade:1c,Jafarzade:1d,Nampalliwar:1b,Salvatore:1a,Zeng:2021dlj}. 

The BH shadow in $F(R)$ gravity was studied in \cite{Dastan:1b}. In Ref. \cite{Sun:1b}, the authors investigated the geometric shadow shape of BHs in $F(R)$ gravity with nonlinear electrodynamics. The study of shadow and constraining parameters was carried out using EHT observational data for BHs in $F(R, T)$ gravity \cite{Hazarika:djb}, and Hu-Sawicki $F(R)$ gravity \cite{Karmakar:969}. In this work, we consider a BH in $F(R)$-ModMax gravity and explore its optical properties, as well as constrain the parameters with the help of the EHT data of M87*.

The manuscript is structured as follows:  in section \ref{Sec1}, we briefly review the charged BHs in $F(R)-$ModMax theory and examine the admissible region to have a physical BH solution. In section \ref{BHS}, we analyze the shadow behavior of this BH solution and show how the shadow size is affected by changes in model parameters. In subsection \ref{EHT}, we compare the resulting shadows to the one detected by the EHT collaboration and study the constraints on the parameters of the model. We then calculate the energy emission rate and explore the effect of the $F(R)-$ModMax parameters on the emission of particles in subsection \ref{EER}. The deflection angle of light around this BH and the effective role of the model's parameters on this optical quantity will be discussed in section \ref{LDA}. Finally, the main results of our study will be provided in section \ref{conclusion}.

\section{Charged BHs in $F(R)-$ModMax Theory}
\label{Sec1}
In this section, we first review the charged AdS BH solution in the $F(R)-$ModMax gravity. In such a theory of gravity,  the $F(R)-$ModMax field  (as the source of matter) is coupled with $F(R)$ gravity. The action of $F(R)-$ModMax theory is given by \cite{Panah:1b}
\begin{equation}\label{b1}
	\mathcal{S}_{F(R)}=\frac{1}{16 \pi}\int_{\partial \mathcal{M}}d^{4}x\sqrt{-g}[F(R)-4\mathcal{L}],
\end{equation}
where $F(R)=R+f(R)$ in which $R$ denotes the scalar curvature and $f (R)$ is a function of Ricci scalar curvature. $\mathcal{L}$ refers to the Lagrangian of $F(R)-$ModMax defined as
\begin{equation}\label{b2}
	\mathcal{L}=\frac{1}{2}\left(\mathcal{R}\cosh\gamma-\sqrt{\mathcal{R}^{2}+\mathcal{P}^{2}}\sinh\gamma\right),
\end{equation}
in which $\gamma$ is a dimensionless parameter known as $F(R)-$ModMax theory parameter. $\mathcal{R}$ and $\mathcal{P}$ are, respectively, a true scalar and a pseudoscalar, which are written in the following forms
\begin{align}\label{b3}
	&\mathcal{R}=\frac{\mathcal{F}}{2}\nonumber\\
	&\mathcal{P}=\frac{\tilde{\mathcal{F}}}{2},
\end{align}
where $\mathcal{F}=F_{\mu\nu}F^{\mu \nu}$ ($F_{\mu\nu}=\partial_{\mu}A_{\nu}-\partial_{\nu}A_{\mu}$, in which $A_{\mu}$ is the gauge potential) denotes electromagnetic tensor and $
\widetilde{\mathcal{F}}=F_{\mu \nu }\widetilde{F}^{\mu \nu }$,
where $\widetilde{F}^{\mu \nu }=\frac{1}{2}\epsilon _{\mu \nu }^{~~~\rho
\lambda }F_{\rho \lambda }$. Considering $\mathcal{P}=0$, the equations of motion are obtained as 
\begin{align}\label{b4}
	&R_{\mu\nu}(1+f_{R})-\frac{g_{\mu\nu}F(R)}{2}+(g_{\mu\nu} \bigtriangledown^{2}-\bigtriangledown_{\mu}\bigtriangledown_{\nu})f_{R}=8\pi T_{\mu\nu}\nonumber\\
	&\partial_{\mu}(\sqrt{-g}\tilde{E}^{\mu\nu})=0,
\end{align}
where $f_{R}=\frac{df(R)}{dR}$ and the energy-momentum tensor $T_{\mu\nu}$ is given by
\begin{equation}\label{b5}
	4\pi T^{\mu\nu}=(F^{\mu\sigma}F^{\nu}_{\sigma}e^{-\gamma})-e^{-\gamma}\mathcal{R}g^{\mu\nu},
\end{equation}
then $\tilde{E}_{\mu\nu}$ is defined as
\begin{equation}\label{b6}
	\tilde{E}_{\mu\nu}=\frac{\partial \mathcal{L}}{\partial F^{\mu \nu}}=2(\mathcal{L}_{\mathcal{R}}F_{\mu\nu}),
\end{equation}
in which $\mathcal{L}_{\mathcal{R}}=\frac{\partial \mathcal{L}}{\partial \mathcal{R}}$. For the electrically charged case, the ModMox field equation $\eqref{b4}$  can be reduced to
\begin{equation}\label{b7}
	\partial_{\mu}(\sqrt{-g}e^{-\gamma}F^{\mu \nu})=0.
\end{equation}
Now, one can consider a static spherically symmetric spacetime as
\begin{equation}\label{b8}
	ds^{2}=-h(r)dt^{2}+\frac{dr^{2}}{h(r)}+r^{2}(d\theta^{2}+\sin^{2}\theta \,\ d\phi^{2}).
\end{equation}
To obtain the metric function $h(r)$, one can assume the constant scalar curvature of $R = R_{0}=cte$. Then the trace of equation $\eqref{b4}$ becomes
\begin{equation}\label{b9}
	R_{0}(1+f_{R_{0}})-2(R_{0}+f(R_{0}))=0,
\end{equation}
where $f_{R_{0}}=f_{R_{\arrowvert_{R=R_{0}}}}$. Solving Eq. $\eqref{b9}$ in terms of $R_{0}$ leads to
\begin{equation}\label{b10}
	R_{0}=\frac{2f(R_{0})}{f_{R_{0}}-1}.
\end{equation}
Now inserting Eq. $\eqref{b10}$ into Eq. $\eqref{b4}$, the equations of motion can be rewritten as
\begin{equation}\label{b11}
	R_{\mu\nu}(1+f_{R_{0}})-\frac{g_{\mu \nu}}{4}R_{0}(1+f_{R_{0}})=8\pi T_{\mu\nu}.
\end{equation}
In this step, one needs to obtain a radial electric field which is possible by considering a consistent gauge potential  as $A_{\mu}=u(r)\delta^{t}_{\mu}$. Substituting this gauge potential in Eq. (\ref{b7}),  one finds
\begin{equation}\label{b11a}
	u(r)=-\frac{q}{r},
\end{equation} 
in which $q$ denotes an integration constant related to the electric charge.
With these equations in hand, the metric function $h(r)$ can be obtained as \cite{Panah:1b}
\begin{equation}\label{b12}
	h(r)=1-\frac{m_{0}}{r}-\frac{R_{0}r^{2}}{12}+\frac{q^{2}e^{-\gamma}}{(1+f_{R_{0}})r^{2}},
\end{equation}
where $m_{0}$ is an integration constant. Setting $ f_{R_{0}}=0 $, $  R_{0}=4\Lambda$, and $ \gamma=0 $, the solution (\ref{b12}) becomes Reissner-Nordstr\"om-(A)dS black hole.
\begin{figure}[!htb]
\centering
\subfloat[]{
        \includegraphics[width=0.33\textwidth]{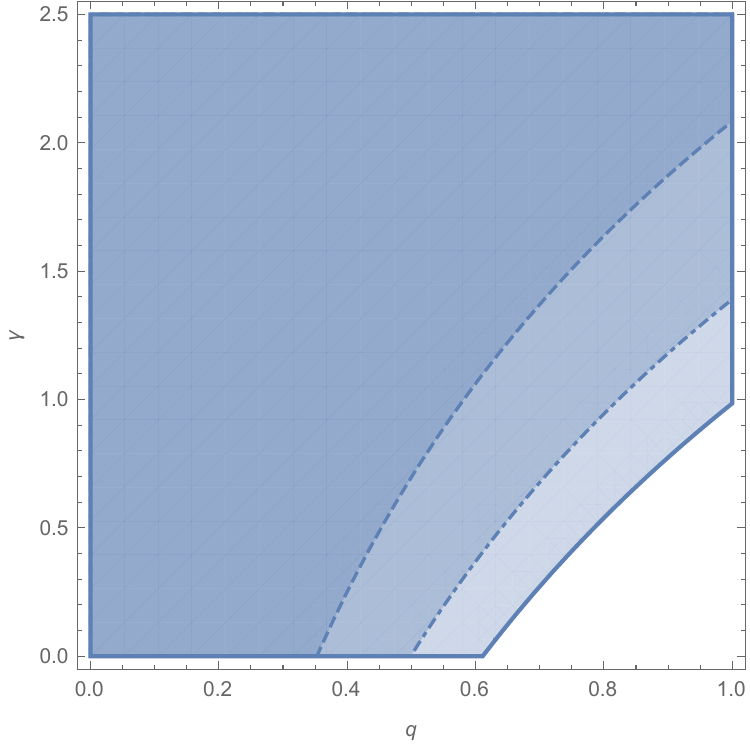}}
\subfloat[]{
     \includegraphics[width=0.33\textwidth]{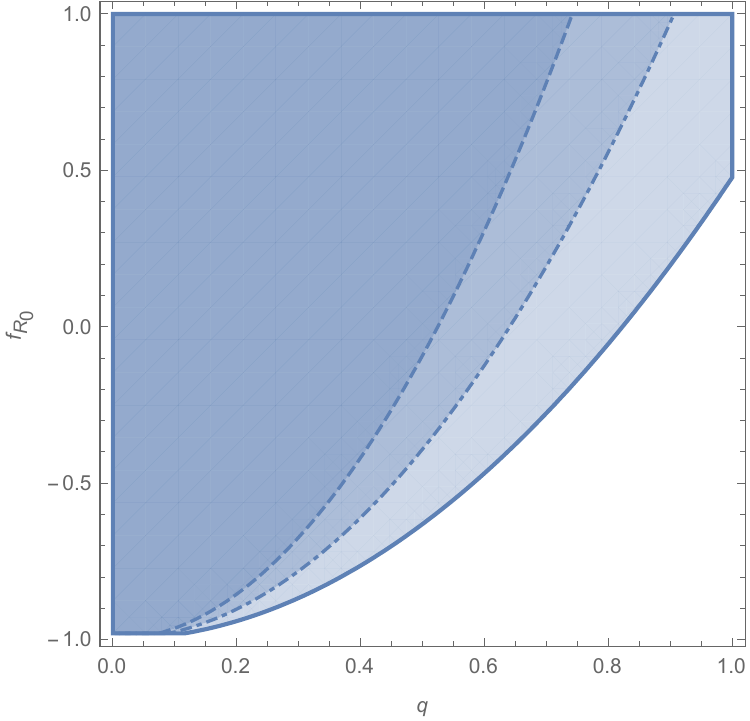}}
     \newline
\subfloat[]{
        \includegraphics[width=0.325\textwidth]{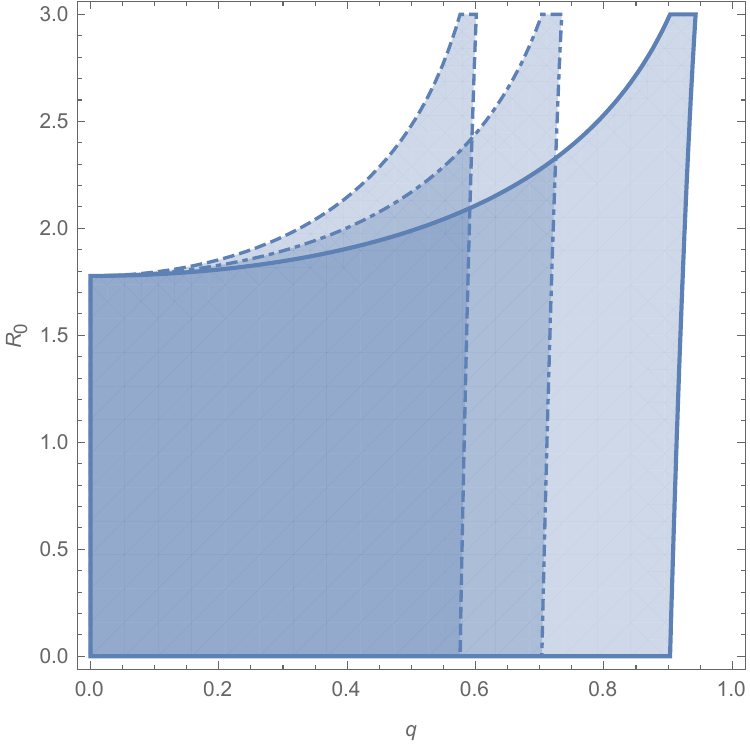}}
\subfloat[]{
        \includegraphics[width=0.33\textwidth]{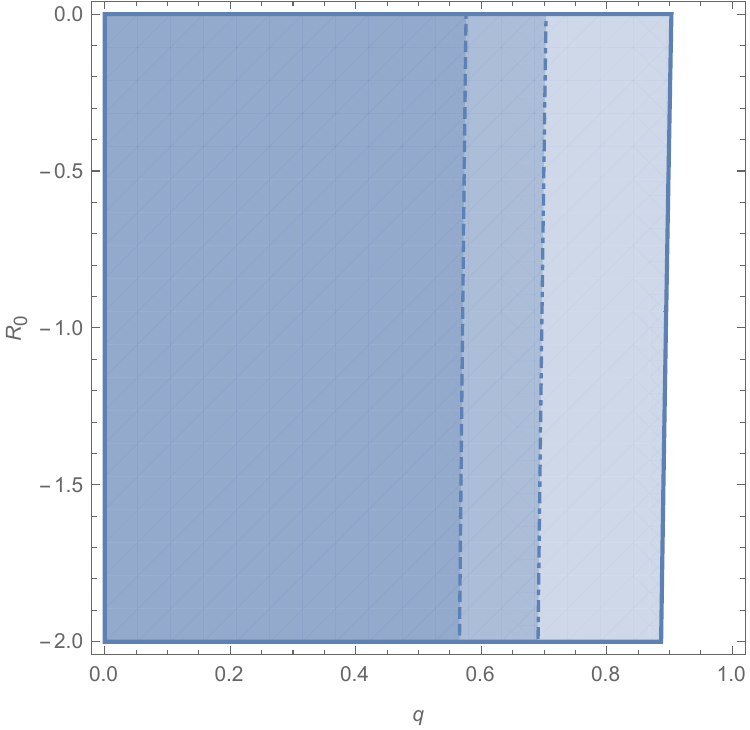}}        
\newline
\caption{The admissible region (denoted by shaded areas) is displayed in (a): $ \gamma -q $ plane for $ R_{0}=-0.2$ and $f_{R_{0}}=-0.5$ (dashed Curve), $f_{R_{0}}=0$ (dashdotted curve) and $f_{R_{0}}=0.5$ (continuous curve); (b): $f_{R_{0}} - q$ plane for $ R_{0}=-0.2$ and $\gamma=0.1$ (dashed curve), $\gamma=0.5$ (dashdotted curve) and $\gamma =1$ (continuous curve); (c): $R_{0} - q$ plane for $ f_{R_{0}}=0.2$ and $\gamma=0.1$ (dashed curve), $\gamma=0.5$ (dashdotted curve) and $\gamma =1$ (continuous curve); (d): $R_{0} - q$ plane for $ f_{R_{0}}=0.2$ and $\gamma=0.1$ (dashed curve), $\gamma=0.5$ (dashdotted curve) and $\gamma =1$ (continuous curve).}
\label{Fig1}
\end{figure}

Before studying the optical features of this solution, we would like to explore the allowed regions of parameters in which a physical solution exists. Fig. $(\ref{Fig1})$ shows the admissible parameter space (denoted by colored areas) to have a BH solution. From this figure, it can be seen that the admissible parameter space is highly affected by the values of the parameters of the model. According to Fig. $\ref{Fig1}$(a), $f_{R_{0}}$ parameter increases the region related to the physical solution. Looking at  Figs. $\ref{Fig1}$(b), $\ref{Fig1}$(c), and $\ref{Fig1}$(d) one finds that a similar discussion can be considered for the effect of $ \gamma $ on the allowed region. A remarkable point regarding the $R_{0}$  effect is that 
there is an upper bound for positive values of $R_{0}$ to have a BH solution (see Fig. $\ref{Fig1}$(c) for more details). 

\section{Optical properties}
In this section, we would like to study the optical properties of black
holes, such as the shadow geometrical shape and the energy emission rate, and discuss how the electric charge and the $F(R)-$ModMax gravity parameters
affect the size of the shadow and the emission of particles around the BHs. To do so, we employ the geodesic equation to calculate the radius of the innermost circular
orbit for a photon in the BH spacetime.
\subsection{Black hole shadow}
\label{BHS}
The BH shadow is an optical feature
of the BH when a bright distant source is behind it or when the BH is surrounded by an optically bright accretion disk. In other words, a distant observer sees a BH as a two-dimensional dark zone in the sky on the background of other bright sources, this dark zone is referred to as a "BH shadow". The shadow of a BH is created by the sphere of photons surrounding it.  The photon ring is the result of light deflection or gravitational lensing. The strong gravitational field near the event horizon of a BH affects light paths and produces spherical light rings. For spherically symmetric BHs, the boundary of the shadow is a perfect circle, while for rotating ones, the BH shadow is no longer circular but rather flattened on one side, as a consequence of the "dragging" of lightlike geodesics by the BH. 

\begin{figure}[!htb]
\centering
\includegraphics[width=0.46\textwidth]{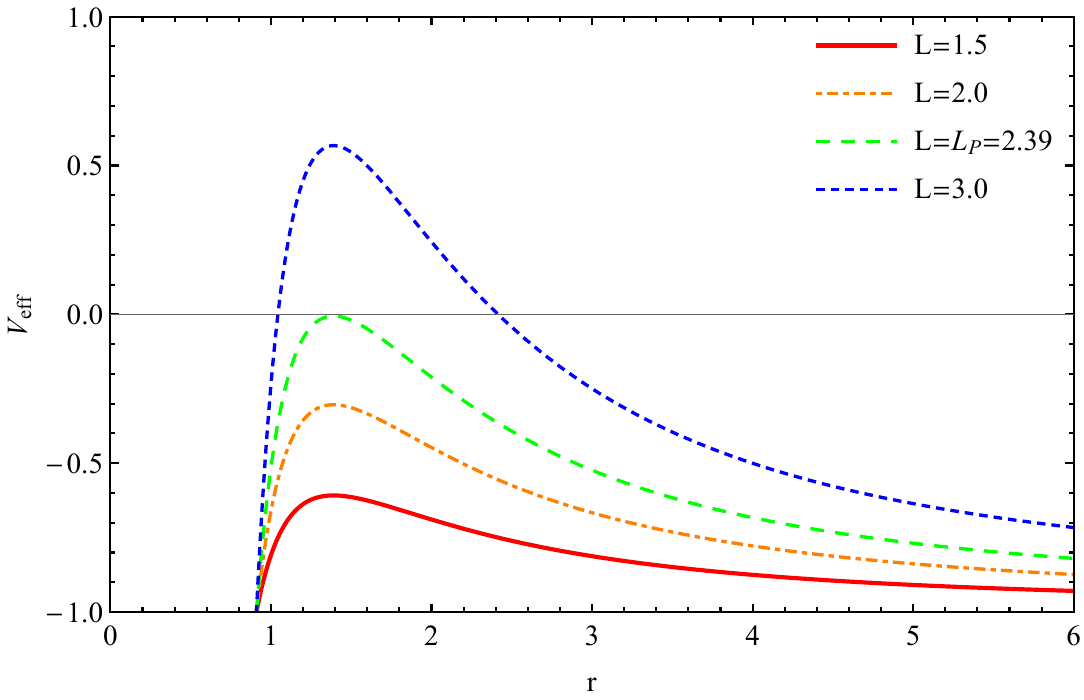}        
\newline
\caption{The behavior of effective potential $V_\text{eff} (r)$ for $E =
M = 1$, $ q=0.5 $, $ R_{0}=-0.1 $, $ \gamma =1 $ and $f_{R_{0}}=0.2$  with different
values of $ L $.}
\label{Fig0}
\end{figure}

For our purpose, we consider a general form of a static spherically symmetric spacetime as follows \cite{Karmakar:dj}
\begin{equation}\label{sh1}
g_{\mu \nu}	dx^{\mu} dx^{\nu}=-A(r)dt^{2}+B(r)dr^{2}+D(r)(d\theta^{2}+\sin^{2}\theta d\phi^{2}),
\end{equation}
in which $A(r)$, $B(r)$ and $D(r)$ can be arbitrary positive function of $r$.
For the case of a spherically symmetric and static spacetime metric, the Lagrangian can be expressed as \cite{Karmakar:dj,Karmakar:djb}
\begin{equation}\label{a}
\mathfrak{L}(x,\dot{x})=\frac{1}{2}g_{\mu \nu}\dot{x}^{\mu}\dot{x}^{\nu}	=\dfrac{1}{2}[-A(r)\dot{t}^{2}+B(r)\dot{r}^{2}+D(r)(\dot{\theta}^{2}+\sin^{2}\theta \dot{\phi}^{2})],
\end{equation}
where the overdot indicates the derivative with respect to an affine parameter $\tau$. 
Due to the spherically symmetric property of the BH, we consider trajectories of photons
on the equatorial plane with $\theta = \dfrac{\pi}{2}$.
The Euler-Lagrange equations are given by
\begin{equation}\label{a0}
\frac{d}{d\tau}(\frac{\partial \mathfrak{L}}{\partial \dot{x}^{\mu}})-\frac{\partial \mathfrak{L}}{\partial x^{\mu}}=0.
\end{equation}
Solving the above equations (for $\mu=0$ and $\mu=3$), the two constants of motion for the  motion of photons are obtained
\begin{align}\label{a1}
&E=A(r) \dot{t}\nonumber\\
&L=D(r)\dot{\phi},
\end{align}
where $E$ and $L$ represent the energy and angular momentum of the photon, respectively. Using the normalization condition for the 4-velocity $u^\mu u_\mu=0,$ for photons leads to
\begin{equation}\label{a2}
A(r)\dot{t}^{2}+B(r) \dot{r}^{2}+D(r) \dot{\phi}^{2}=0.
\end{equation}
Then by incorporating Eq. $\eqref{a1}$ to Eq. (\ref{a2}), one can obtain a complete description of the radial
motion as
\begin{equation}\label{a3}
\dot{r}^{2}=\frac{1}{B(r)}\left[ \frac{E^{2}}{A(r)}-\frac{L^{2}}{D(r)}\right] .
\end{equation}
The term on the right side of Eq. (\ref{a3}) is defined as the effective potential
\begin{equation}\label{a4}
\dot{r}^{2}=V_\text{eff}.
\end{equation}
In Fig. \ref{Fig0}, the behavior of the effective potential is plotted for different values of $L$, indicating that the maximum of the potential increases with an increase in $ L $. According to the fact that $ \dot{r}^{2}>0 $, it is expected that the effective potential satisfies $ V_\text{eff}\leq 0 $. From this, it can be seen that photon trajectories appear only for negative effective potential. Therefore, an incoming photon from infinity falls into the BH for smaller values of $ L $. In contrast, photons with larger values of $ L $ are reflected before falling into the BH. An interesting  phenomenon is related to the critical
angular momentum $ L=L_{P} $ ($max(V_\text{eff})=0$). In such a situation, the photon loses its radial velocity and acceleration at $r=r_{Max}$ (the case of $r=r_{Max}$ is the same photon orbit denoted by $r=r_{ph}$) and orbits the BH due to its non-vanishing transverse velocity. Therefore, the photon orbits are circular and unstable, associated with the maximum of the effective potential. To obtain this maximum value, the following conditions must be considered simultaneously:
\begin{equation}
\label{v12}
V_\text{eff} \arrowvert_{r=r_{ph}} = 0, \,\,\ \frac{d V_\text{eff}}{d r}\arrowvert_{r=r_\text{ph}}=0.
\end{equation}
Solving the first equation yields the critical angular momentum ($L_{P}$), while the second equation determines the photon orbit radius ($ r_\text{ph} $). Inserting Eq. (\ref{a3}) into Eq. (\ref{a4}) and solving $ V_{eff}^{\prime}(r_{ph})=0 $ gives
\begin{equation}
 \frac{d V_\text{eff}}{d r}\arrowvert_{r=r_\text{ph}}=
D^{\prime} (r_\text{ph})A(r_{p})-A^{\prime}(r_{p})D(r_{p})=0,
\end{equation}
which provides $r_\text{ph}$ as 
\begin{equation}\label{rph0}
 r_\text{ph}=\frac{3}{4}\left(m_{0} +\sqrt{m_{0}^{2}-\frac{32 q^{2} }{9 e^{\gamma}\left(1+ f_{R_{0}}\right) }}\right). 
\end{equation}

From the initial conditions,  the size of BH shadow can be obtained as
\begin{equation}\label{rsh}
r_\text{sh}=\frac{r_\text{ph}}{\sqrt{A(r_\text{ph})}}.
\end{equation}

	Using celestial coordinates $X$ and $Y$, the shadow's apparent shape can be obtained as
	\begin{align}
		& X=\lim_{r_{0}\to \infty}\left(-r_{0}^{2} \sin\theta_{0}\frac{d\phi}{dr}\arrowvert_{(r_{0},\theta_{0})}\right)\\
		&Y=\lim_{r_{0}\to \infty}\left(r_{0}^{2}\frac{d \theta}{dr}\arrowvert_{(r_{0},\theta_{0})}\right),
	\end{align}
	where $\theta_{0}$ represents the angular position of the observer with respect to the plane of the BH. For our solution, the celestial coordinates $X$ and $Y$ are determined as
\begin{equation}
X=-\frac{r_\text{sh}}{\sqrt{1+\frac{R_{0}}{12}r_\text{sh}^{2}}}  , \,\,\,\,\,\,\,\,\ Y=0.
\end{equation}

\begin{figure}[!htb]
\centering
\subfloat[$ R_{0}=-0.1 $, $ \gamma =1 $ and $f_{R_{0}}=0.2$]{
        \includegraphics[width=0.33\textwidth]{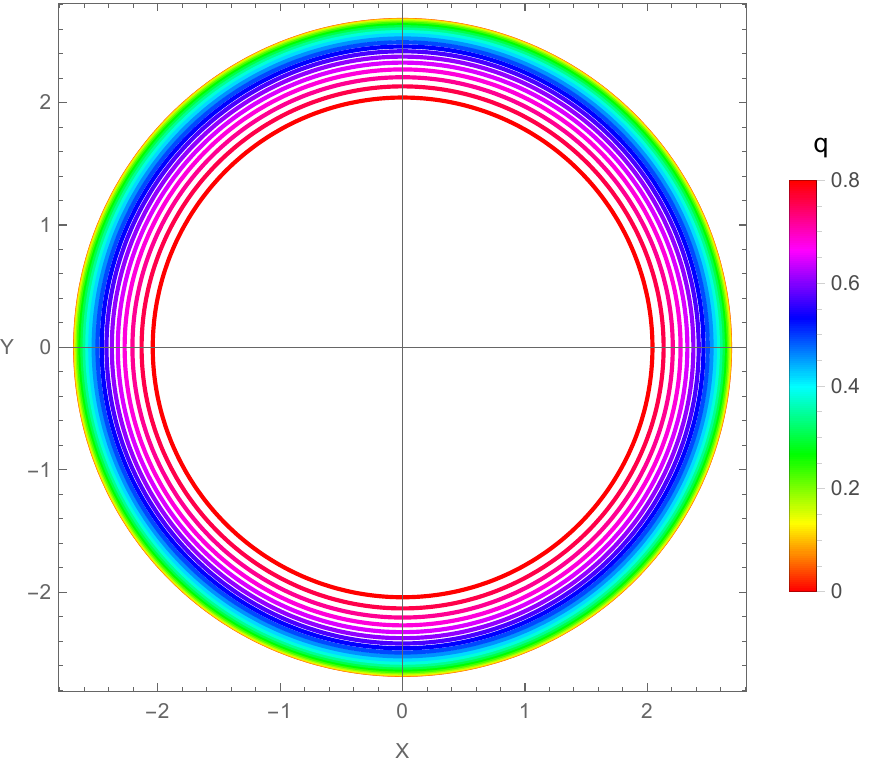}}
\subfloat[$ R_{0}=-0.1 $, $ q =0.6 $ and $f_{R_{0}}=0.2$]{
     \includegraphics[width=0.33\textwidth]{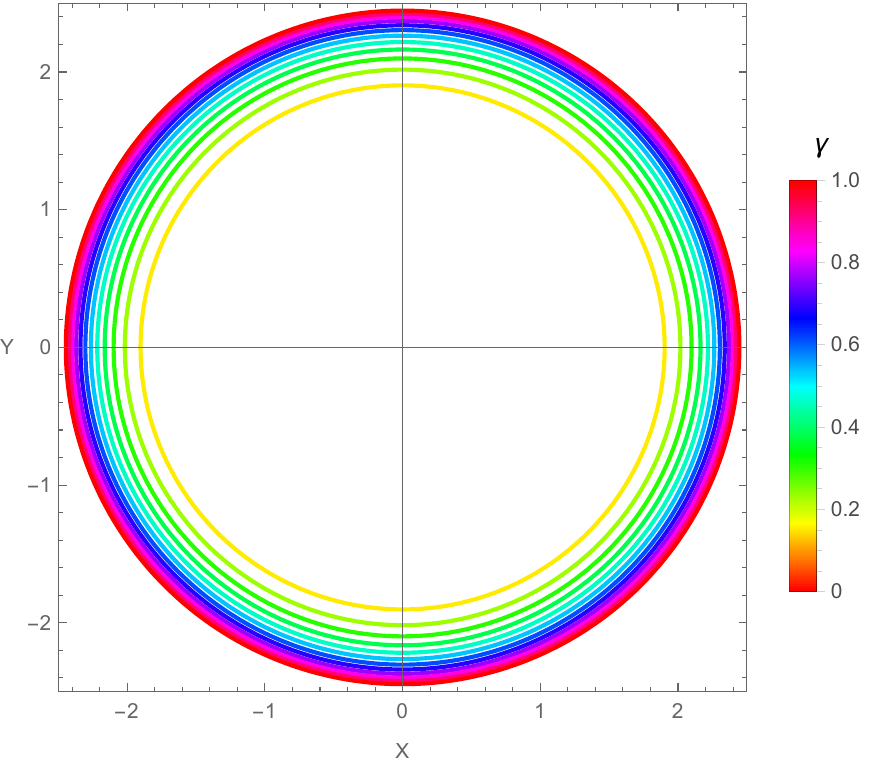}}
\subfloat[$ R_{0}=-0.1 $, $ q =0.6 $ and $\gamma =1$]{
     \includegraphics[width=0.33\textwidth]{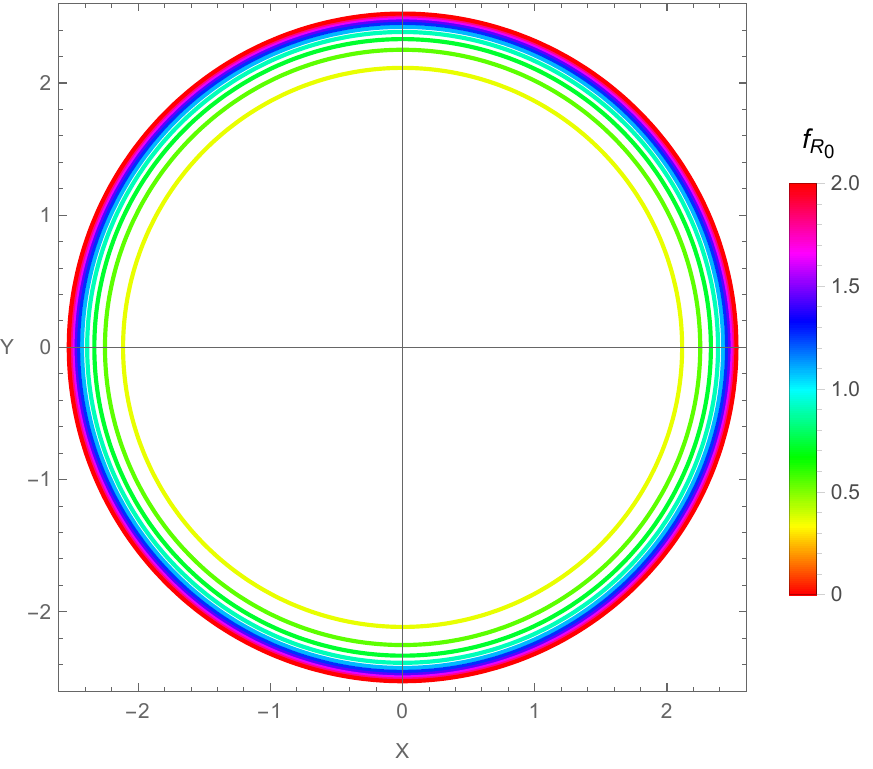}}
     \newline
\subfloat[$ \gamma=1 $, $ q =0.6 $ and $f_{R_{0}}=0.2$]{
        \includegraphics[width=0.33\textwidth]{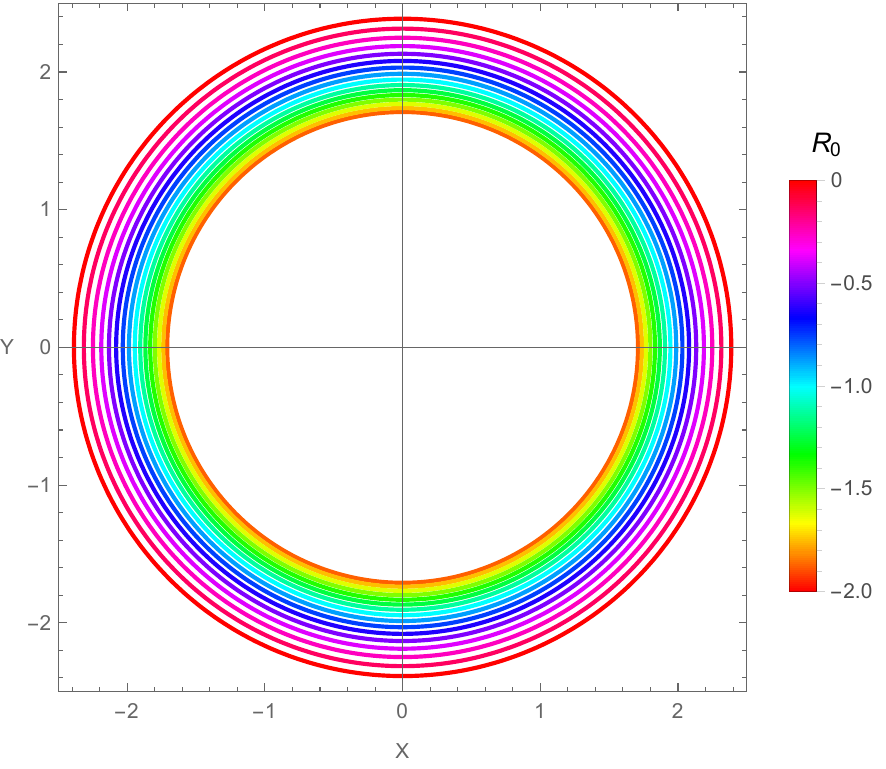}}
\subfloat[$ \gamma=1 $, $ q =0.6 $ and $f_{R_{0}}=0.2$]{
        \includegraphics[width=0.33\textwidth]{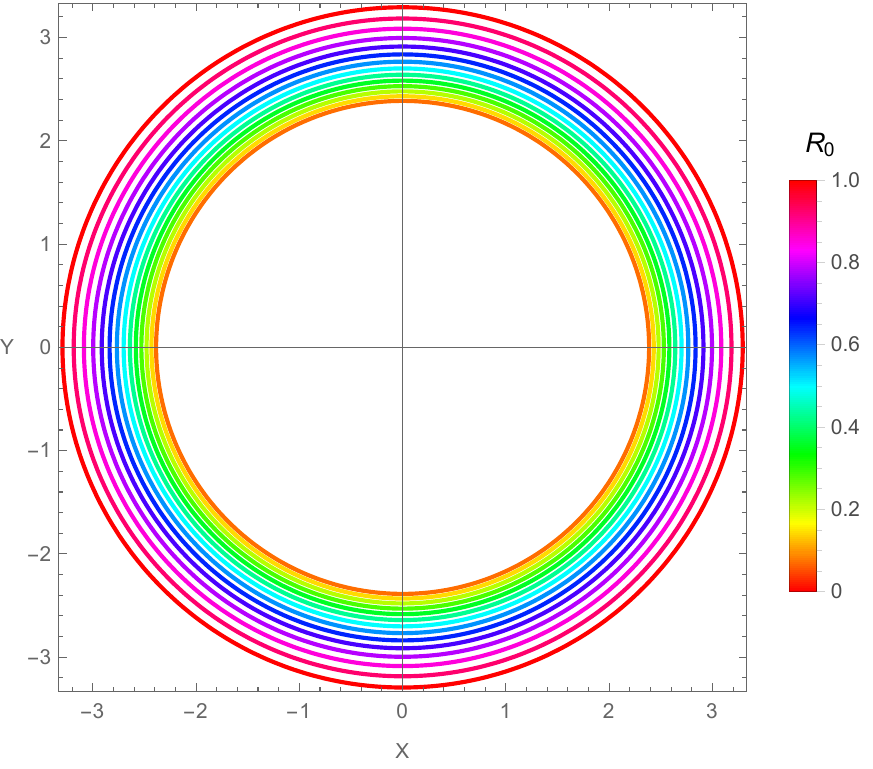}}        
\newline
\caption{The boundary of BH shadow with changing $ q $, $ \gamma $, $ f_{R_{0}} $ and $ R_{0} $.}
\label{Fig2}
\end{figure}

To show how the shadow radius is affected by the parameters of the model, we have plotted Fig. \ref{Fig2}. Fig. \ref{Fig2}(a) represents the influence of the electric charge on the shadow radius, revealing the fact that the shadow size shrinks in the presence of the electric field. This feature was observed separately in various articles. For instance, in Ref. \cite{Zhu:100}, the authors studied the optical properties of BHs in the Einstein-\AE ther theory and showed that increasing the electric charge leads to a decrease in the shadow radius. In the perspective of nonlinear electrodynamics, a study on the optical properties of rotating BHs is presented in \cite{Zubair:948,Raza:4410}, which demonstrates that the BH shadow decreases with increasing charge parameter. As an another instance, the BH shadow was investigated under three different models of nonlinear electrodynamics: Born-Infeld, Euler-Heisenberg, and Modified Maxwell in Ref. \cite{Guzman:002} and was illustrated that the electric charge has a decreasing effect on the shadow size for all three cases\footnote{The reason why the shadow radius decreases with the increase in the charge parameter is that charged black holes are generally smaller than neutral black holes. This is also evident from the fact that Schwarzschild BH has the radius $2M$ while the Reissner-Nordstr\"{o}m extremal BH has the radius $M$. Thus increasing electric charge parameter of BH basically shrinks the BH and hence the respective shadow size, as confirmed by the reported literature.}.  The effect of ModMax parameter $ \gamma $ has been illustrated in Fig. \ref{Fig2}(b), indicating that increasing this parameter leads to an increase in the radius of shadow which is consistent with the results of Ref. \cite{Sun:1b}, verifying that the nonlinear coupling parameter has an increasing effect on the shadow size. From Fig. \ref{Fig2}(c), one can find the size of shadow
image increases with an increase in $F(R)$ gravity parameter $f_{R_{0}}$ which is similar with what obtained in Ref. \cite{Dastan:1b}. 
The impact of the curvature of the space-time can be seen in Figs. \ref{Fig2}(d) and \ref{Fig2}(e). According to Fig. \ref{Fig2}(d), for negative values of $ R_{0} $ (or in AdS background), the shadow size decreases with an increase in $\vert R_{0}\vert$, while an opposite effect will be observed in dS background or for positive values of $ R_{0} $ (see Fig. \ref{Fig2}(e)). 

\subsection{Observational constraints from the EHT}
\label{EHT}
A possible way to distinguish modified gravity from general relativity is to compare its resulting shadow with observational data. In this subsection, we aim to consider BHs in $F(R)-$ModMax theory as a supermassive BH in M87* and use the EHT observations to further constrain the model parameters. 
Such a study helps us to properly understand the mentioned model and explore its validity.

\begin{figure}[!htb]
\centering
\subfloat[$ \gamma=1 $ and $ f_{R_{0}}=0.2$]{
        \includegraphics[width=0.33\textwidth]{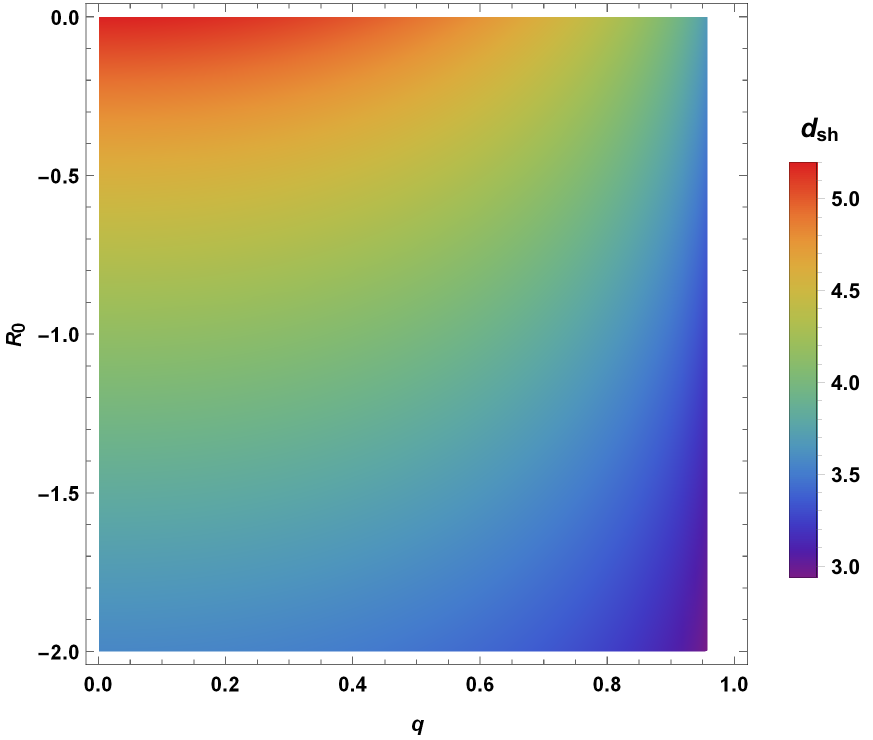}}
\subfloat[$ f_{R_{0}}=0.2$ and $ R_{0}=-0.1 $]{
     \includegraphics[width=0.33\textwidth]{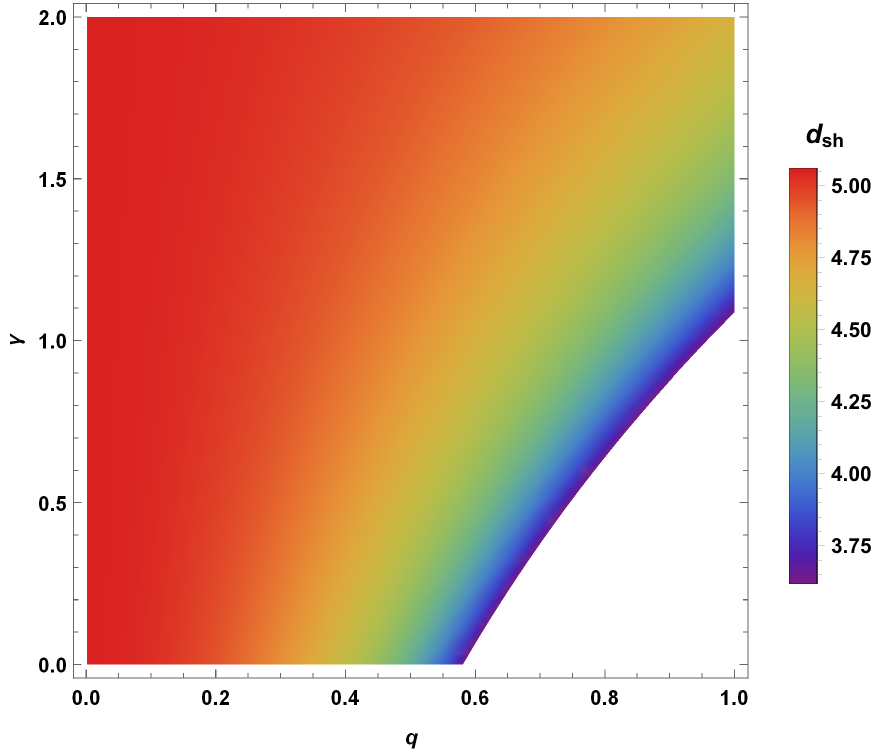}}
\subfloat[$ \gamma=1 $ and $ R_{0}=-0.1$]{
        \includegraphics[width=0.33\textwidth]{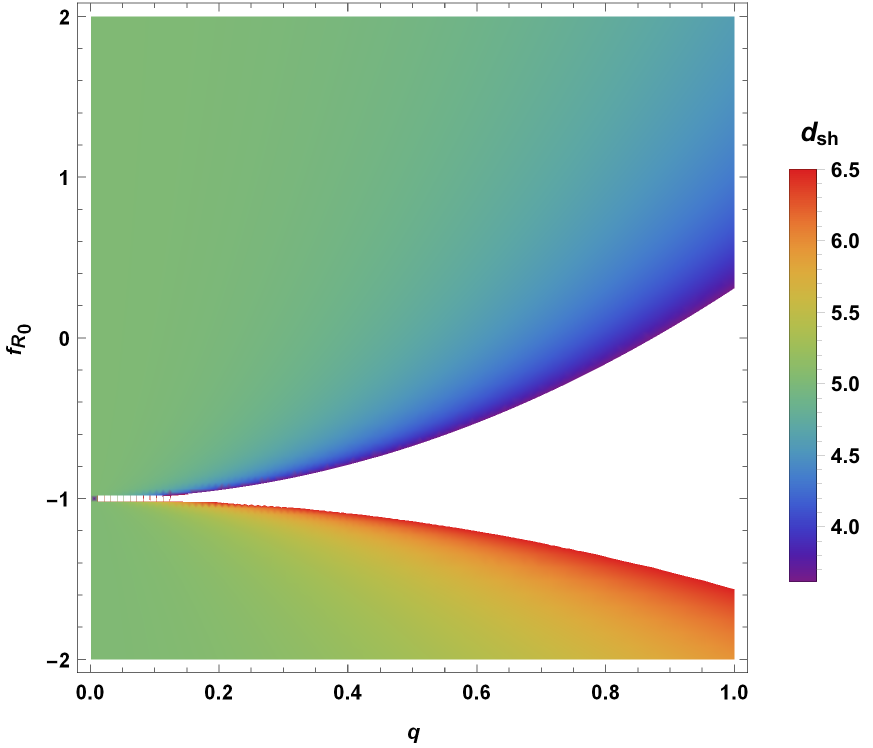}}       
\newline
\caption{The density plots of the shadow diameter which show acceptable regions in agreement with observational data of M87*.}
\label{Fig4}
\end{figure}

\begin{figure}[!htb]
\centering
\subfloat[$ \gamma=1 $ and $ f_{R_{0}}=-1.05$]{
        \includegraphics[width=0.33\textwidth]{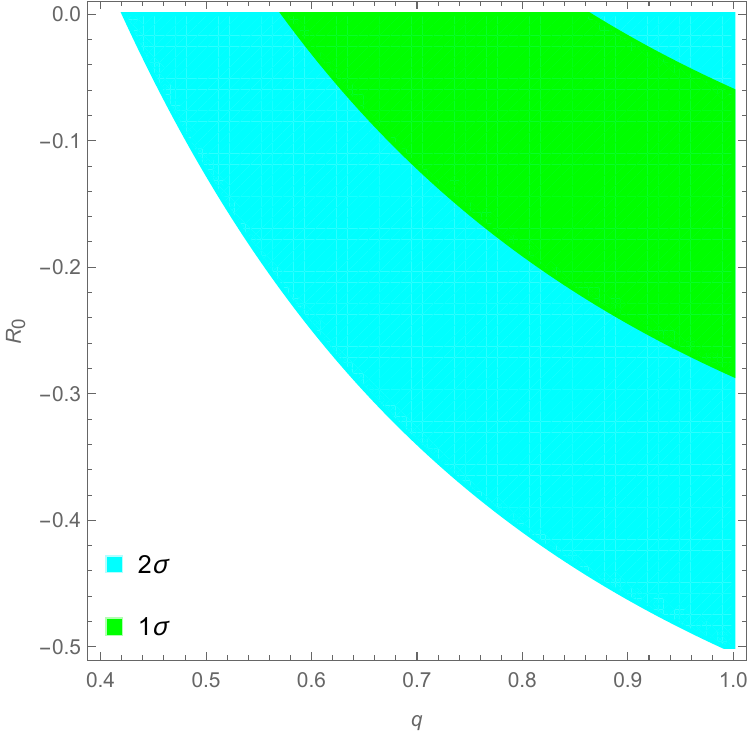}}
\subfloat[$ f_{R_{0}}=-1.05$ and $ R_{0}=-0.1 $]{
     \includegraphics[width=0.33\textwidth]{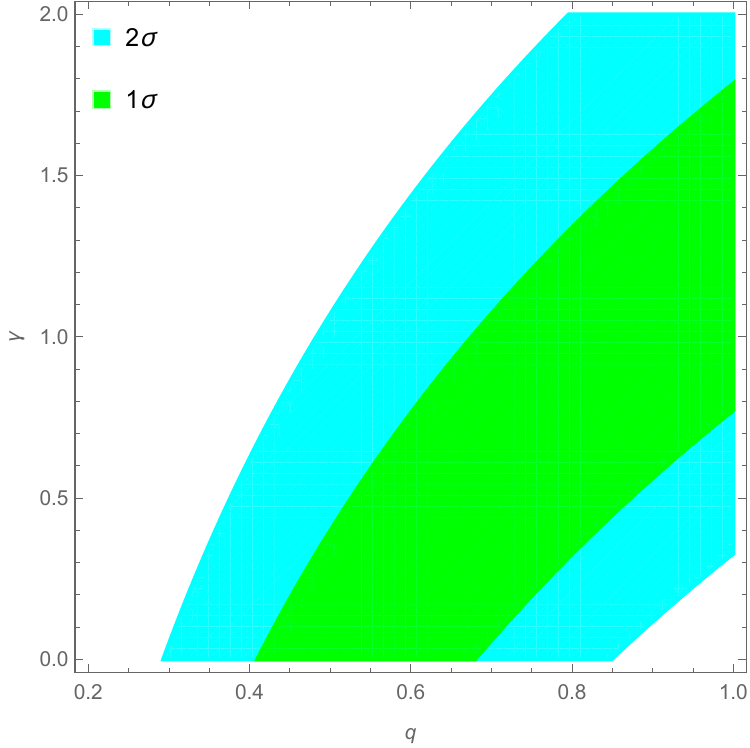}}
\subfloat[$ \gamma=1 $ and $ R_{0}=-0.1$]{
        \includegraphics[width=0.33\textwidth]{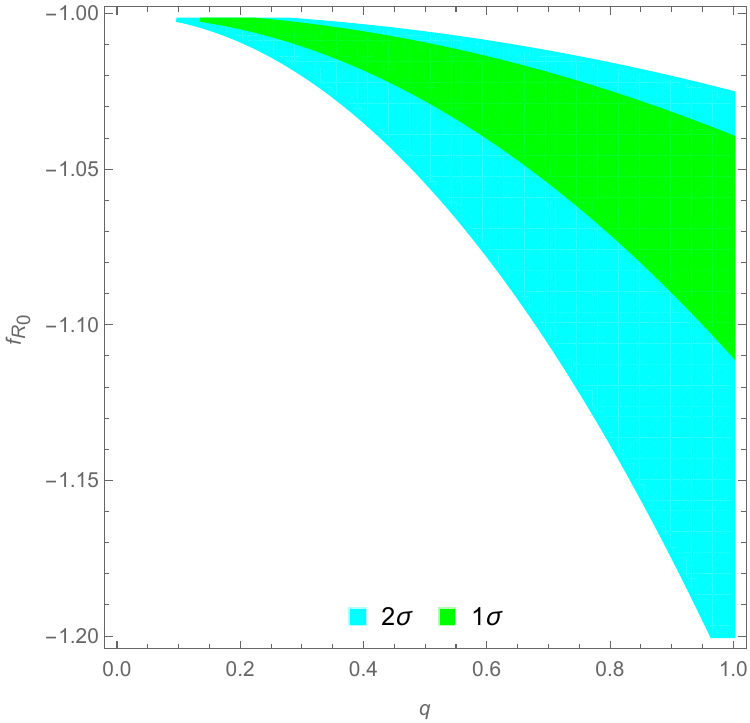}}       
\newline
\subfloat[$ \gamma=1 $ and $ f_{R_{0}}=-1.05$]{
        \includegraphics[width=0.33\textwidth]{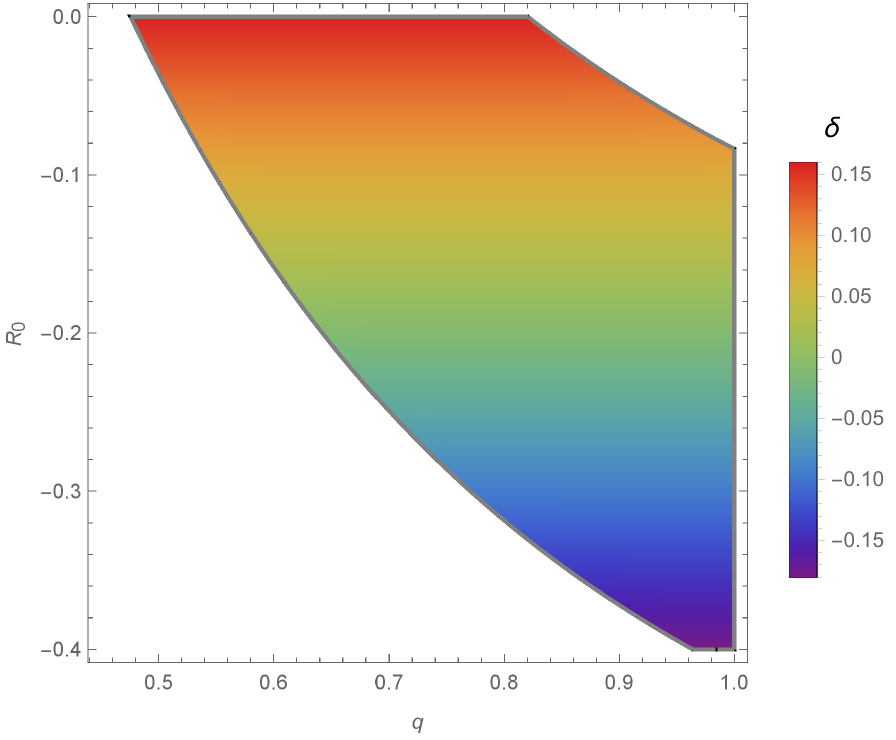}}
\subfloat[$ f_{R_{0}}=-1.05$ and $ R_{0}=-0.1 $]{
     \includegraphics[width=0.33\textwidth]{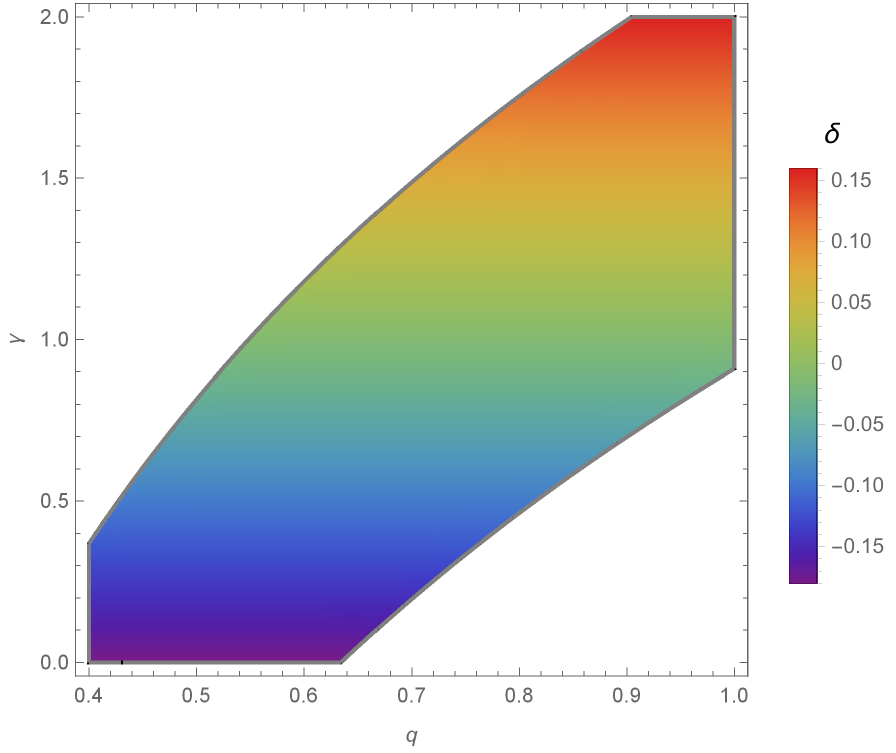}}
\subfloat[$ \gamma=1 $ and $ R_{0}=-0.1$]{
        \includegraphics[width=0.33\textwidth]{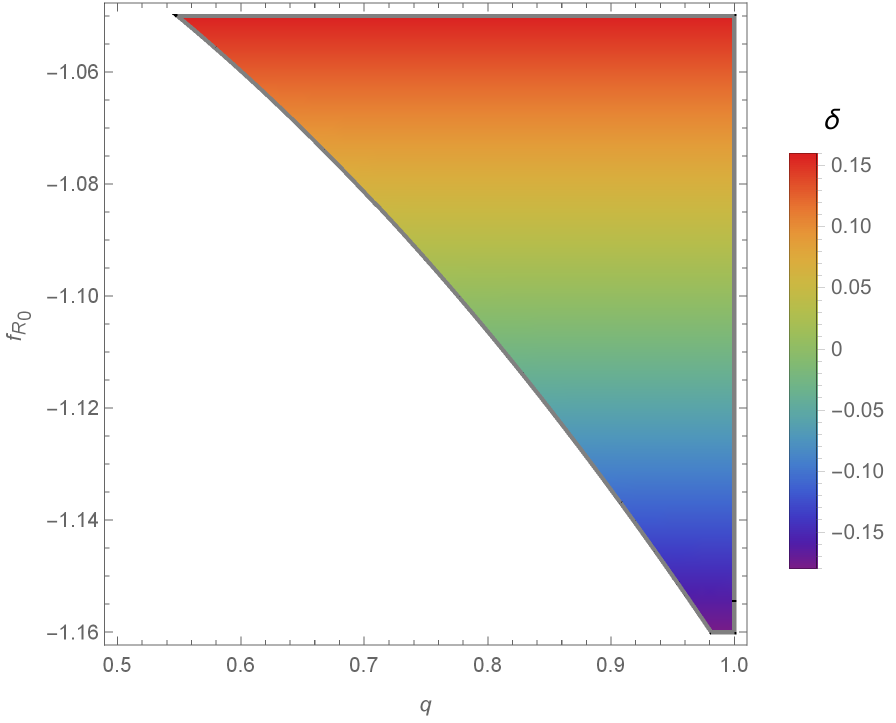}}       
\newline
\caption{\textbf{Top row:} Constraints on the $F(R)-$ModMax gravity's parameters and the electric charge with the EHT observations of M87* for AdS BHs. \textbf{Bottom row:} The shadow diameter deviation
from that of a Schwarzschild BH as a function of
$R_{0}$, $ \gamma $, $ f_{R_{0}} $, and $ q $ for AdS BHs. }
\label{Fig5}
\end{figure}

\begin{figure}[!htb]
\centering
\subfloat[$ \gamma=1 $ and $ f_{R_{0}}=2$]{
        \includegraphics[width=0.33\textwidth]{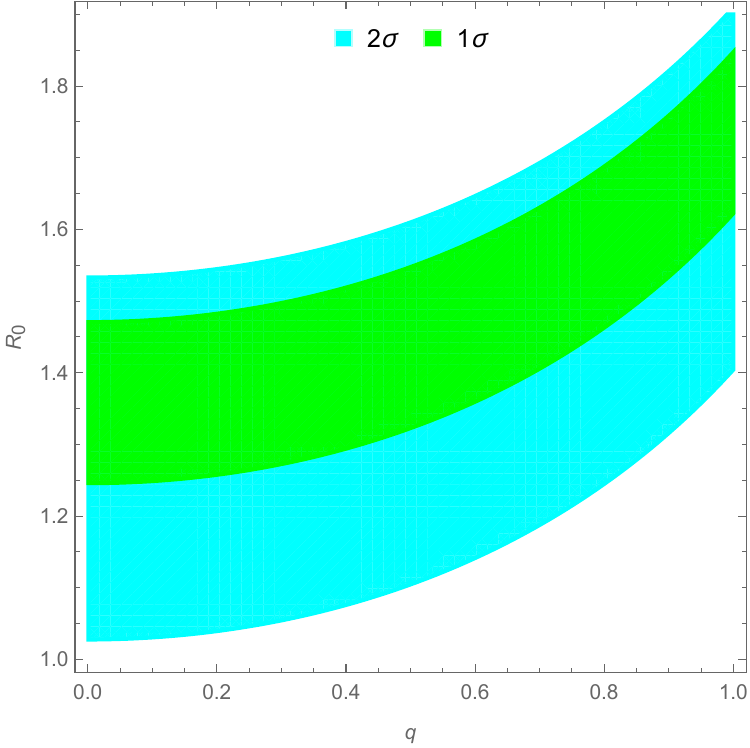}}
\subfloat[$ f_{R_{0}}=2$ and $ R_{0}=1.5 $]{
     \includegraphics[width=0.33\textwidth]{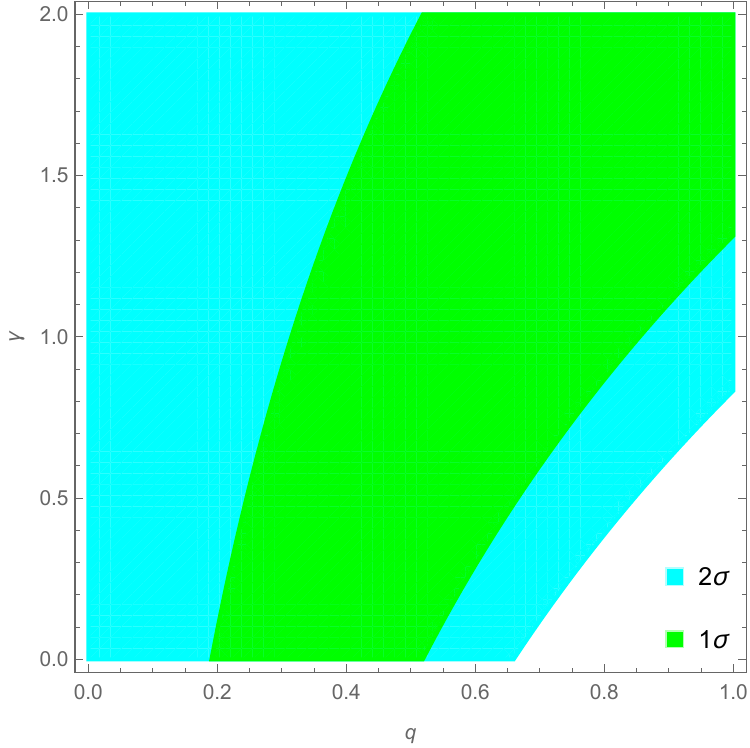}}
\subfloat[$ \gamma=1 $ and $ R_{0}=1.5$]{
        \includegraphics[width=0.33\textwidth]{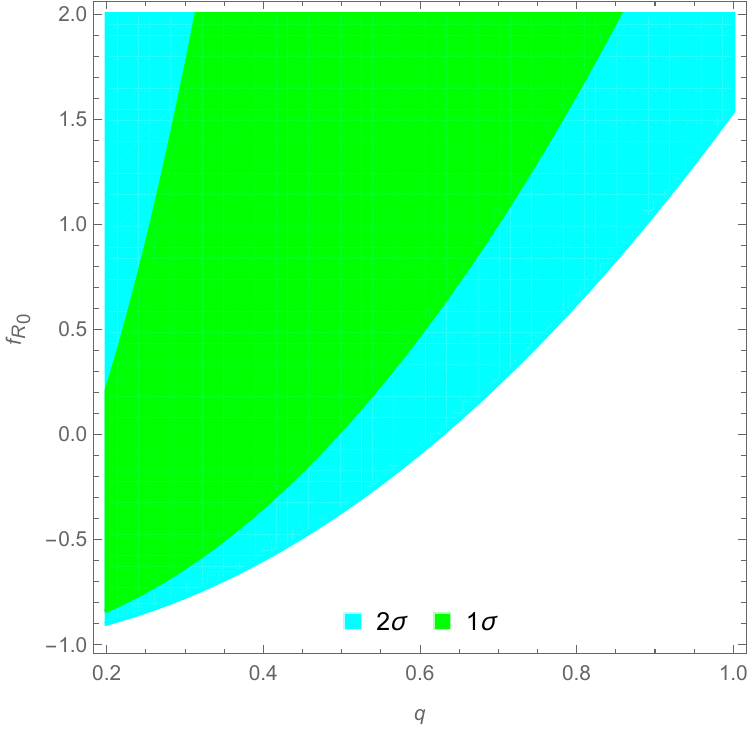}}       
\newline
\subfloat[$ \gamma=1 $ and $ f_{R_{0}}=2$]{
        \includegraphics[width=0.33\textwidth]{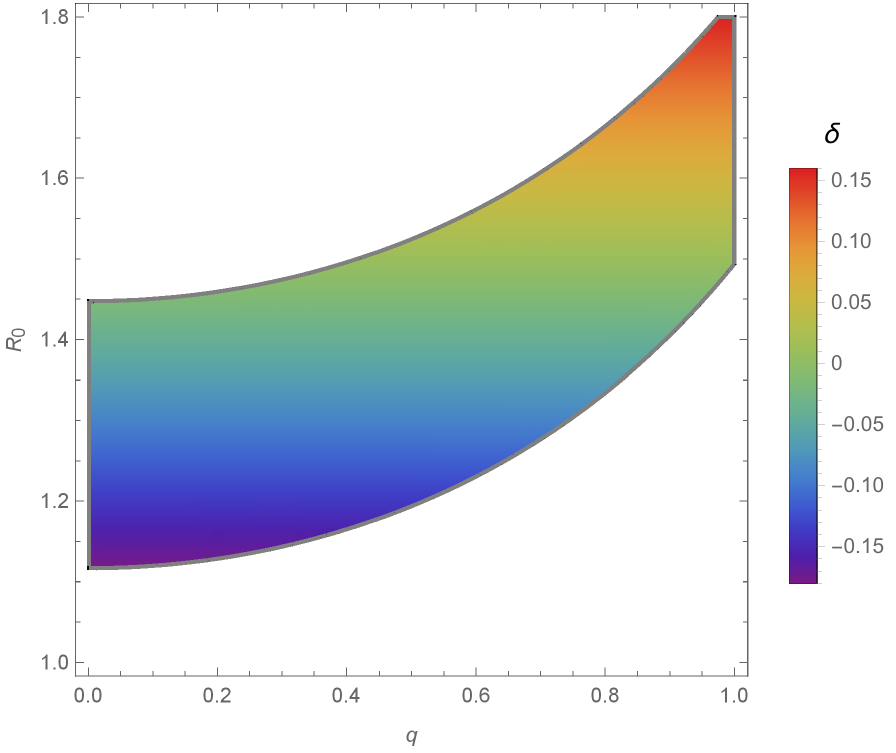}}
\subfloat[$ f_{R_{0}}=2$ and $ R_{0}=1.5 $]{
     \includegraphics[width=0.33\textwidth]{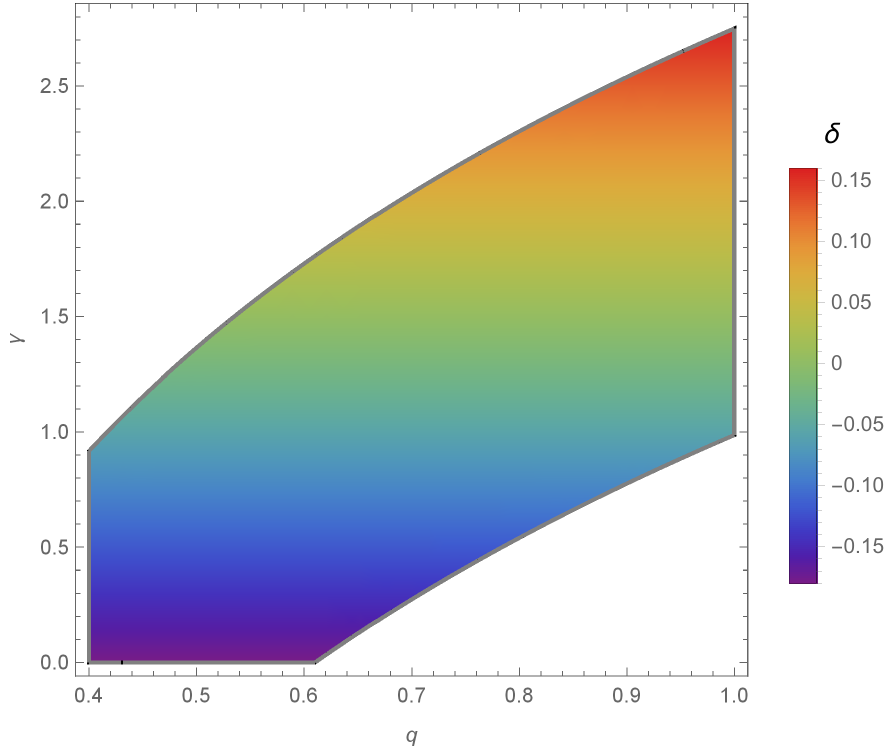}}
\subfloat[$ \gamma=1 $ and $ R_{0}=1.5$]{
        \includegraphics[width=0.33\textwidth]{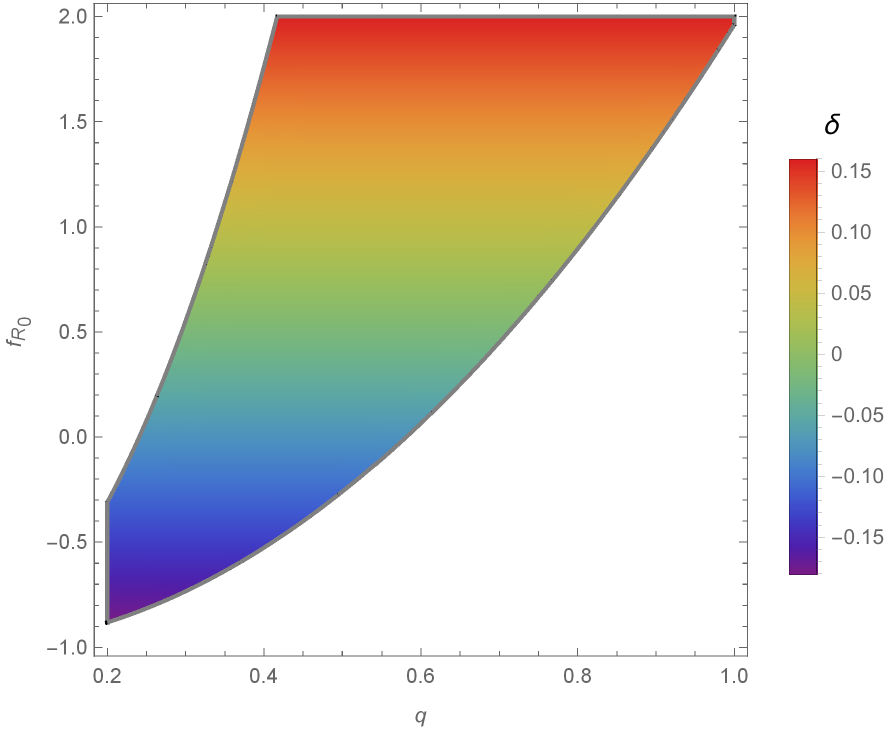}}       
\newline
\caption{\textbf{Top row:} Constraints on the $F(R)-$ModMax gravity's parameters and the electric charge with the EHT observations of M87* for dS BHs. \textbf{Bottom row:} The shadow diameter deviation
from that of a Schwarzschild BH as a function of
$R_{0}$, $ \gamma $, $ f_{R_{0}} $, and $ q $ for dS BHs. }
\label{Fig6}
\end{figure}

According to the obtained results by the EHT collaboration, the supermassive BH at the center of the
galaxy M87* has the following values \cite{Kumar:kg}

\begin{eqnarray}
\theta &=& (42 \pm 3) \mu as,\\
M &=& (6.5 \pm 0.9) \times 10^{9} M_{\odot},\\
\mathbb{D} &=& 16.8^{+0.8}_{-0.7} Mpc,
\label{EqdM87a}
\end{eqnarray}
where $ M $
and $M_{\odot}$ are the mass of the object and Sun respectively. Further, $ \theta $ and $D$ denote the angular diameter
of the shadow and the distance of the M87* from the Earth, respectively.  Using these numbers, the diameter of the shadow in units of mass
can be obtained as \cite{Akiyama:eht}
\begin{equation}
d_{M87^{*}}\equiv \frac{\mathbb{D}\theta}{M}\approx 11.0 \pm 1.5.
\label{EqdM87b}
\end{equation}
According to Eq. (\ref{EqdM87b}), within $ 1\sigma $ uncertainty $9.5 \lesssim d_{M87^{*}}\lesssim 12.5$ whereas within $ 2\sigma $ uncertainty
$8 \lesssim d_{M87^{*}}\lesssim 14$.
To find the allowed regions of the model parameters, we  have plotted Figs. (\ref{Fig4}) and (\ref{Fig5}) (for negative values of $ R_{0} $) and Fig. (\ref{Fig6}) (for positive values of $ R_{0} $). In the uppermost plots in Figs. (\ref{Fig5}) and (\ref{Fig6}), the green-colored area corresponds to the area that is consistent with the observational data within $ 1\sigma $-error. While the cyan-colored area shows the region which is in agreement with EHT data within $ 2\sigma $ uncertainty. From Figs. \ref{Fig4}(a) and \ref{Fig4}(b), it can be seen that $d_{sh}<6$ for positive values of parameter $ f_{R_{0}}$, while for $ f_{R_{0}}<-1$, the shadow diameter can have greater values (see Fig. \ref{Fig4}(c)).  Top panels of Fig. (\ref{Fig5}) verify that the constraint (\ref{EqdM87b}) can be satisfied for $ f_{R_{0}}<-1$. Fig. (\ref{Fig6}) displays the admissible parameter space region for $ R_{0}>0 $ (dS case). From this figure, one finds that a consistent result with EHT data can be observed for $ f_{R_{0}}>-1$. Therefore, $ f_{R_{0}}$ parameter is the main parameter in making the simulated shadow to be consistent with the M87* data. 

Recent results regarding astronomical observations of $M87^{*}$ have estimated the
Schwarzschild shadow deviation $ (\delta) $ which measures the difference
between the model shadow diameter $ (d_{metric}) $ and the Schwarzschild BH
shadow diameter defined as \cite{Akiyama:et}
\begin{equation}\label{d3}
\delta=\frac{d_{metric}}{6\sqrt{3}}-1,
\end{equation}
where $d_{metric} = 2r_{sh}$. According to reported results, the bound of
the measured Schwarzschild deviation is as $-0.18<\delta<0.16$. Clearly, $ \delta $ is positive (negative) if the BH shadow size is greater (smaller) than the Schwarzschild BH of the same mass. Bottom panels of Figs. (\ref{Fig5}) and (\ref{Fig6}) depict the region of parameters which satisfy the mentioned constraint for $\delta$. 
By having a closer look at Fig. (\ref{Fig5}), one can find that for non-zero values of the electric charge and ModMax parameter ($ \vert R_{0} \vert $ and $ \vert f_{R_{0}} \vert $), an AdS BH in $F(R)-$ModMax gravity with the same mass as Schwarzschild, has a greater shadow size compared to the Schwarzschild BH.  While for charged dS BHs, the shadow size gets greater than the Schwarzschild BH
shadow for large values of the parameters.   
 
\subsection{Energy emission}
\label{EER}
\begin{figure}[!htb]
\centering
\subfloat[$ \gamma =1 $, $ R_{0}=-0.1 $ and $ f_{R_{0}}=0.2 $]{
        \includegraphics[width=0.33\textwidth]{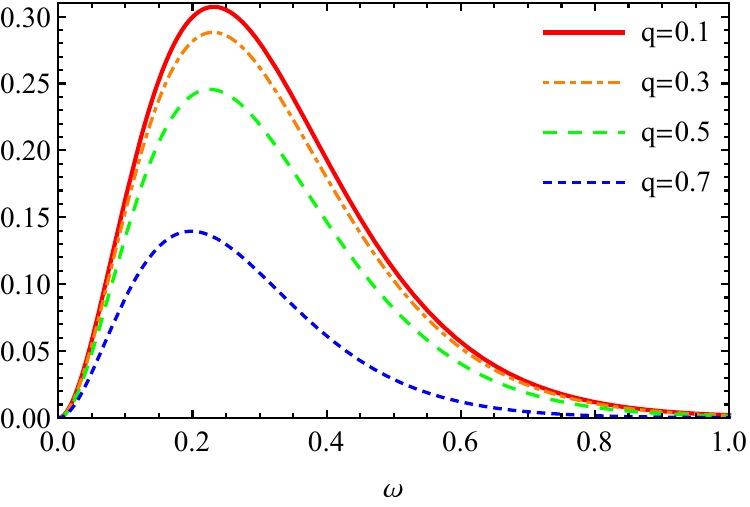}}
\subfloat[$ q =0.6 $, $ R_{0}=-0.1 $ and $ f_{R_{0}}=0.2 $]{
     \includegraphics[width=0.33\textwidth]{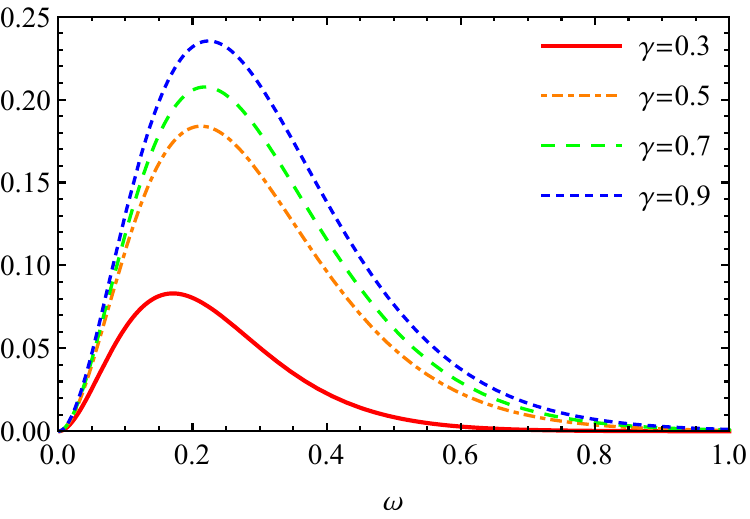}}
\subfloat[$ q =0.6 $ and $ R_{0}=-0.1 $]{
     \includegraphics[width=0.33\textwidth]{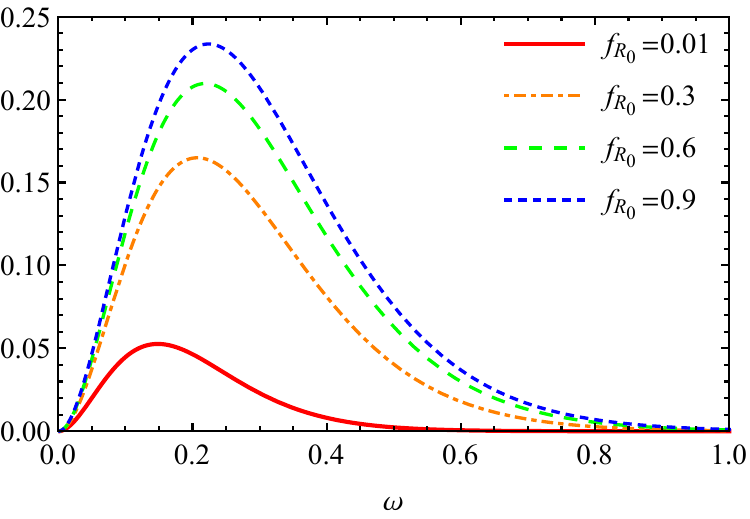}}
     \newline
\subfloat[$ \gamma=1 $, $ q =0.6 $ and $ f_{R_{0}}=0.2 $]{
        \includegraphics[width=0.33\textwidth]{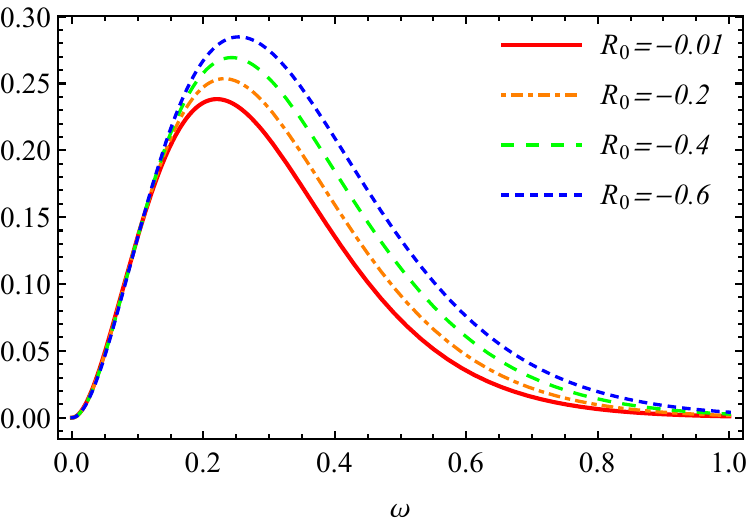}}
\subfloat[$ \gamma=1 $, $ q =0.6 $ and $ f_{R_{0}}=0.2 $]{
        \includegraphics[width=0.33\textwidth]{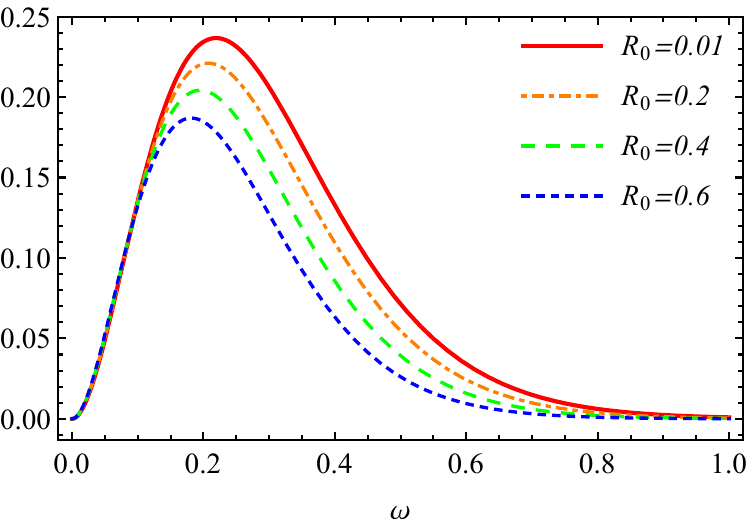}}        
\newline
\caption{The energy emission rates for the corresponding BH, with
$M = 1$, and different values of $ q $, $ \gamma $, $ R_{0} $ and $ f_{R_{0}}=0.2 $.}
\label{Fig3}
\end{figure}

Quantum oscillations in the proximity of BHs cause the generation and annihilation of an excessive number of particles near the horizon. This process leads to the tunneling phenomenon causing pair production with a positive energy particle outside the event horizon and a negative energy antiparticle inside the horizon. Receiving a negative amount of energy reduces the mass of the BH and gradually evaporates it over some time. The energy emission rate is associated with this process, meaning that quantum fluctuations in the interior of BHs are the source of emission energy. It has been shown that the high energy absorption cross-section tends to approach the shadow of the BH for an observer at a very far distance. In addition, in the limit of extremely high energies, the absorption cross-section typically fluctuates until it reaches a constant limiting value $\sigma_{lim}\approx \pi r_{sh}^{2}$ which is approximately equal to the area of the photon sphere. The energy emission rate of a black
hole solution can be expressed as \cite{Wei:L}
\begin{equation}
	\frac{d^{2}E(\omega)}{dt d\omega}=\frac{2 \pi^{2} \omega^{3} r_{sh}^{2}}{\exp^{\frac{\omega}{T}}-1},
\end{equation}
where $\omega$ represents the emission frequency and $T$ denotes the Hawking temperature. The Hawking temperature, which is related to the surface gravity $\kappa$ on the event horizon, is given by \cite{Panah:1b}
\begin{equation}
	T=\frac{\kappa}{2 \pi}=\frac{1}{4\pi r_{h}}-\frac{R_{0}r_{h}}{16\pi}-\frac{q^{2}e^{-\gamma}}{4\pi (1+f_{R_{0}})r_{h}^{3}},
\end{equation}
in which $r_{h}$ is the horizon radius of the BH. Now, we are in a position to explore the effect of $F(R)-$ModMax parameters on the rate of energy emission. 
In Fig. \ref{Fig3}, these energetic aspects are plotted as a function of the emission frequency for different values of $ q $, $ \gamma $, $ f_{R_{0}} $ and $R_{0}$. It can be seen from this figure that there exists a peak of the energy emission rate for the BH which increases/decreases as the BH parameters increase. From Fig. \ref{Fig3}(a), we verify that the electric charge decreases the energy emission, meaning that the evaporation process would be slower for a BH
located in a more powerful electric field. This reveals the fact that neutral BHs have shorter lifetimes compared to their charged counterparts in $F(R)-$ModMax gravity. Regarding the effect of the ModMax parameter, Fig. \ref{Fig3}(b) shows that increasing this parameter leads to increasing energy emission, meaning that the BH  lifetimes will be shorter in the presence of the ModMax nonlinear field. Fig. \ref{Fig3}(c) shows the effect of $F(R)$ gravity's parameter 
$ f_{R_{0}} $ on the emission rate which verifies that the evaporation process would be faster under $F(R)$ gravity. In other words, the BHs in $F(R)$ gravity have a shorter lifetime compared to the BHs in GR. To show the influence of curvature background on the energy emission rate, we have plotted Figs. \ref{Fig3}(d) and \ref{Fig3}(e). According to these two figures, in AdS spacetime (dS spacetime) a BH has a longer lifetime in a low (high) curvature background.

\section{DEFLECTION ANGLE OF LIGHT}
\label{LDA}
Now we are interested in studying the deflection angle of light using the Gauss Bonnet theorem for the $F(R)-$ModMax BH. The Gauss-Bonnet theorem is an important and attractive theorem in the study of surfaces in mathematics and has significant applications in the field of gravitational lensing in modern times. Since we consider BHs in non-asymptotically flat spacetimes, we employ the method mentioned in Refs. \cite{Kottler:an,Ishihara:ys,Kumaran:ao} to obtain the deflection angle of light. In the Gauss-Bonnet theorem, the deflection angle of light is defined as follows 
\begin{equation}
\alpha=\psi_{R}-\psi_{S}+\phi_{RS},
\end{equation}
where $\psi_{R}$ and $\psi_{S}$ are the angles of the optical beam of the lens, which are measured with respect to the observer and the source, respectively.
 $\phi_{OS}$ is the coordinate of the separation angle between the observer and the source, i.e. $\phi_{RS}\equiv \phi_{R}-\phi_{S}$. 
In the previous section, we were able to obtain the constants of motion for the metric by writing the Lagrangian. These constants include the energy $E=A(r)\dot{t}$ and angular momentum of the photon, i.e. $L=D(r)\dot{\phi}$, which were obtained on the equatorial plane. In the field of gravitational lensing, an important parameter known as the impact parameter $b$ is defined based on these two parameters as follows
\begin{equation}
	b=\frac{L}{E}=\frac{D(r)}{A(r)}\frac{d\phi}{dt}.
\end{equation}

 According to the metric (\ref{sh1}), the  orbital equation of a photon in the spherically symmetric spacetime can be obtained by introducing the inverse radial coordinate $r=\frac{1}{u}$ as
\begin{equation}
	\left(\frac{du}{d\phi}\right)^{2}=\frac{D(r)^{2}u^{2}}{A(r)B(r)} \left[\frac{1}{b^{2}}-\frac{A(r)}{D(r)}\right].
\end{equation}

The orbit equation for the light ray in a static and spherically symmetric spacetime
is in a general form as
\begin{equation}
\left(\frac{du}{d\phi}\right)^{2}=F(u),
\end{equation}
where $ F(u) $ is defined as
\begin{equation}
F(u)=-u^{2}A(u)+\frac{1}{b^{2}}.
\end{equation}

Since $\psi$ represents the angle between the radial component of the tangent vector and the radial vector, i.e. the light angle, it can be defined by
\begin{equation}
\cos\psi=\frac{b}{D(r)}\frac{d\phi}{dt}.
\end{equation}

This leads to
\begin{equation}
	\sin\psi=\sqrt{\frac{b^{2}A(r)}{D(r)}}.
	\label{Psi}
\end{equation}

 Also for $\phi_{OS}$, one has
\begin{equation}
\phi_{OS}=2\int_{0}^{u_{0}}\frac{du}{F(u)}.
\end{equation}

Consequently, when we consider the source and the receiver to be separated by a finite distance, the deflection angle can be represented as
\begin{equation}
\alpha=\psi_{O}-\psi_{S}+\int_{u_{S}}^{u_{0}}\frac{du}{F(u)}+\int_{u_{O}}^{u_{0}}\frac{du}{F(u)}.
\end{equation}

\subsection{Light deflection in $F(R)-$ModMax gravity}

Now we use the mentioned equations to calculate the deflection angle of light by the corresponding BHs in the
weak-field approximation. For our case, the orbit equation is 
\begin{equation}
F(u)=\frac{1}{b^{2}}+\frac{R_{0}}{12}-u(\phi)^{2}+m_{0}u(\phi)^{3}-\mathcal{A}u(\phi)^{4},
\end{equation}
and the solution for the photon orbit is found to be
\begin{equation}
u(\phi)=\frac{\sin \phi }{b}+\frac{m_{0}(1+\cos ^{2}\phi )}{2b^{2}}+\frac{1}{24} b R_{0}\sin \phi +\frac{\left( 3\phi \cos \phi -3\sin \phi +\sin^{3} \phi\right) \mathcal{A}}{4b^{3}},
\label{phi1}
\end{equation}
in which $ \mathcal{A}=\frac{q^{2}e^{ -\gamma}}{1+f_{R_{0}}} $. Using Eq. (\ref{Psi}), $\Psi_{R} $ - $\Psi_{S} $ is expanded in terms of $ m_{0} $, $ R_{0} $ and other parameters as
\begin{eqnarray}
\Psi_{R}-\Psi_{S} &=&\left(\sin ^{-1}(bu_{R})+\sin ^{-1}(bu_{S}) -\pi\right) -\frac{ b m_{0}}{2}\left(\frac{u_{R}^{2}}{\sqrt{1-b^{2}u_{R}^{2}}} +\frac{u_{S}^{2}}{\sqrt{1-b^{2}u_{S}^{2}}}\right)\\ \nonumber
&-&\frac{ b R_{0}}{24}\left(\frac{u_{R}^{-1}}{\sqrt{1-b^{2}u_{R}^{2}}} +\frac{u_{S}^{-1}}{\sqrt{1-b^{2}u_{S}^{2}}}\right)-\frac{ b m_{0}R_{0}}{48}\left(\frac{1}{\sqrt{1-b^{2}u_{R}^{2}}} +\frac{1}{\sqrt{1-b^{2}u_{S}^{2}}}\right)\\ \nonumber
&+&\frac{ b \mathcal{A}}{2}\left(\frac{u_{R}^{3}}{\sqrt{1-b^{2}u_{R}^{2}}} +\frac{u_{S}^{3}}{\sqrt{1-b^{2}u_{S}^{2}}}\right)+\frac{m_{0}R_{0} b^{3} }{48}\left(\frac{u_{R}^{2}}{\left( 1-b^{2}u_{S}^{2}\right)^{\frac{3}{2}}} +\frac{u_{S}^{2}}{\left( 1-b^{2}u_{S}^{2}\right)^{\frac{3}{2}} }\right)\\ \nonumber
&+&\mathcal{O}\left( m_{0}^{2},R_{0}^{2},\mathcal{A}^{2},m_{0}\mathcal{A},R_{0}\mathcal{A}\right).
\label{PsiRS}
\end{eqnarray}

From Eq.(\ref{phi1}), $ \phi_{RS} $ is obtained as
\begin{eqnarray}
\phi_{RS} &=&-\left(\sin ^{-1}(bu_{R})+\sin ^{-1}(bu_{S}) -\pi\right)
+\frac{3\mathcal{A}\left( \sin ^{-1}(bu_{R})+\sin ^{-1}(bu_{S})\right)}{4b^{2}}\\ \nonumber
&+&
\frac{ m_{0}}{2b}\left(\frac{2-b^{2}u_{R}^{2}}{\sqrt{1-b^{2}u_{R}^{2}}} +\frac{2-b^{2}u_{S}^{2}}{\sqrt{1-b^{2}u_{S}^{2}}}\right)+\frac{R_{0} b^{4}-18\mathcal{A} }{24b}\left(\frac{u_{R}}{\sqrt{1-b^{2}u_{R}^{2}}} +\frac{u_{S}}{\sqrt{1-b^{2}u_{S}^{2}}}\right)
\\ \nonumber
&+&
\frac{ b \mathcal{A}}{4}\left(\frac{u_{R}^{3}}{\sqrt{1-b^{2}u_{R}^{2}}} +\frac{u_{S}^{3}}{\sqrt{1-b^{2}u_{S}^{2}}}\right)
+\frac{m_{0}R_{0}}{48}\left(\frac{2-3b^{2}u_{R}^{2}}{\left( 1-b^{2}u_{S}^{2}\right)^{\frac{3}{2}}} +\frac{2-3b^{2}u_{S}^{2}}{\left( 1-b^{2}u_{S}^{2}\right)^{\frac{3}{2}} }\right)\\ \nonumber
&+&\mathcal{O}\left( m_{0}^{2},R_{0}^{2},\mathcal{A}^{2},m_{0}\mathcal{A},R_{0}\mathcal{A}\right). 
\label{phiRS}
\end{eqnarray}

Employing Eqs. (\ref{PsiRS}) and (\ref{phiRS}), we obtain the correct deflection angle of light as
\begin{eqnarray}
\alpha &=&\frac{3\mathcal{A}\left( \sin ^{-1}(bu_{R})+\sin ^{-1}(bu_{S})\right)}{4b^{2}}+\frac{m_{0}}{b}\left(\sqrt{1-b^{2}u_{R}^{2}} +\sqrt{1-b^{2}u_{S}^{2}}\right)-\frac{b R_{0}}{24}\left(\frac{u_{R}^{-1}}{\sqrt{1-b^{2}u_{R}^{2}}} +\frac{u_{S}^{-1}}{\sqrt{1-b^{2}u_{S}^{2}}}\right)\\ \nonumber
&+&\frac{b m_{0} R_{0}}{48}\left(\frac{1}{\sqrt{1-b^{2}u_{R}^{2}}} +\frac{1}{\sqrt{1-b^{2}u_{S}^{2}}}\right)
+\frac{3 b \mathcal{A}}{4}\left(\frac{u_{R}^{3}}{\sqrt{1-b^{2}u_{R}^{2}}} +\frac{u_{S}^{3}}{\sqrt{1-b^{2}u_{S}^{2}}}\right)-\frac{3 \mathcal{A}}{4b}\left(\frac{u_{R}}{\sqrt{1-b^{2}u_{R}^{2}}} +\frac{u_{S}}{\sqrt{1-b^{2}u_{S}^{2}}}\right). 
\label{alpha}
\end{eqnarray}

\begin{figure}[!htb]
\centering
\subfloat[$ \gamma =1 $, $ R_{0}=-0.1 $ and $ f_{R_{0}}=0.2 $]{
        \includegraphics[width=0.33\textwidth]{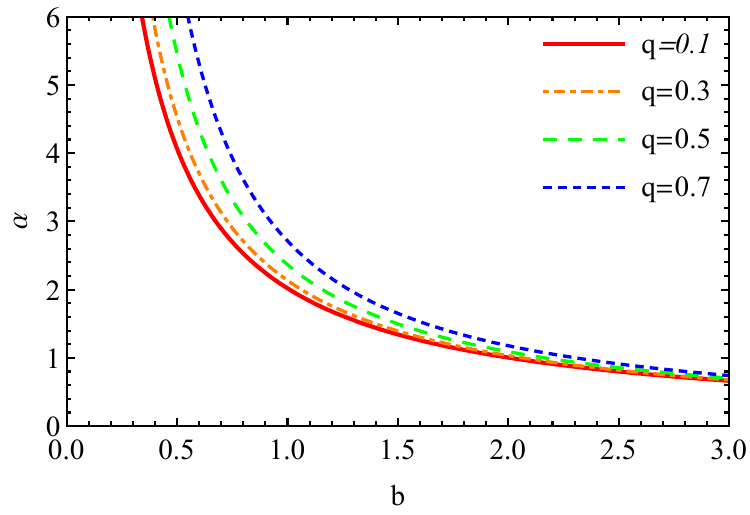}}
\subfloat[$ q =0.6 $, $ R_{0}=-0.1 $ and $ f_{R_{0}}=0.2 $]{
     \includegraphics[width=0.33\textwidth]{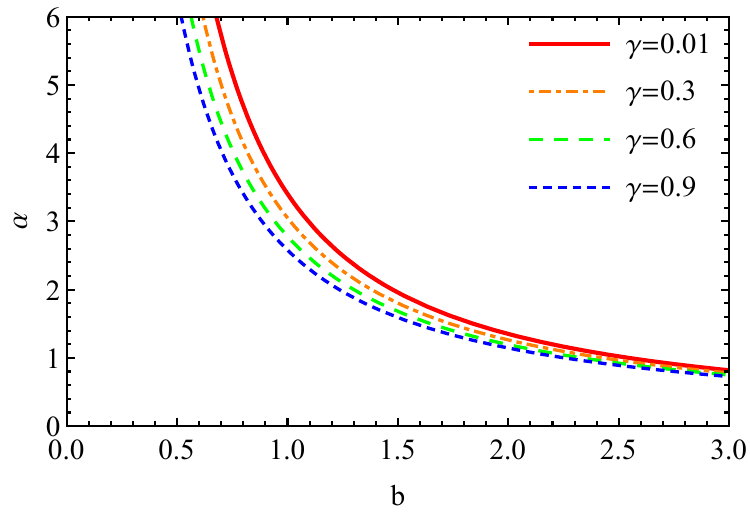}}
\subfloat[$ q =0.6 $, $ R_{0}=-0.1 $ and $ \gamma =1 $]{
     \includegraphics[width=0.33\textwidth]{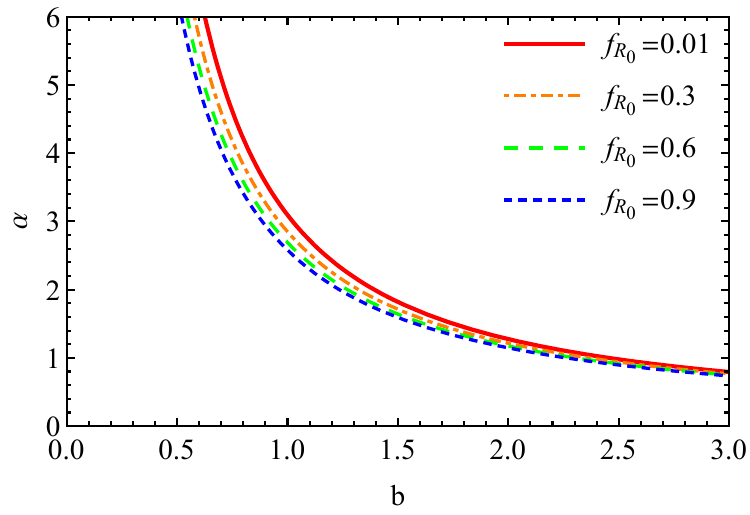}}
     \newline
\subfloat[$ \gamma=1 $, $ q =0.6 $  and $ f_{R_{0}}=0.2 $]{
        \includegraphics[width=0.33\textwidth]{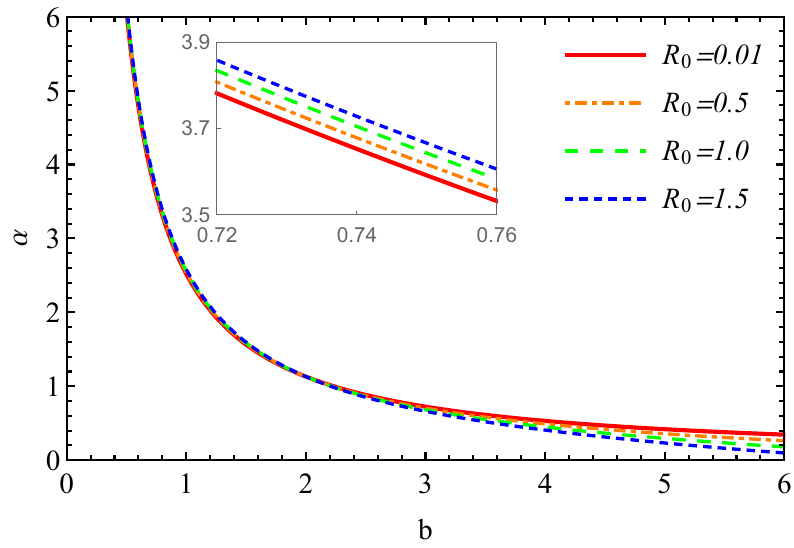}}
\subfloat[$ \gamma=1 $, $ q =0.6 $  and $ f_{R_{0}}=0.2 $]{
        \includegraphics[width=0.33\textwidth]{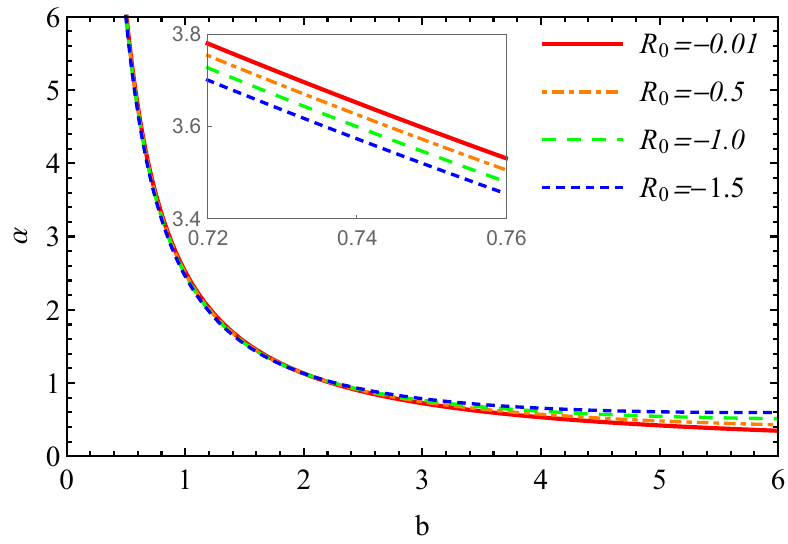}}        
\newline
\caption{Deflection angle vs impact parameter $  b$ for 
$M = 1$ and different values of $ q $, $ \gamma $, $ R_{0} $  and $ f_{R_{0}}=0.2 $.}
\label{Fig7}
\end{figure}

Considering $ u_{S}\rightarrow 0 $ and $ u_{R}\rightarrow 0 $, the deflection angle can be obtained for the infinite distance limit. The behavior of $  \alpha$ concerning the impact parameter $ b $ is illustrated in figure \ref{Fig7}. It can be seen from Fig. \ref{Fig7}(a) that the deflection angle increases as the electric charge increases, indicating that light rays deviate more from the straight path in the presence of the electric field. Fig. \ref{Fig7}(b) shows
the effect of $\gamma$ parameter on the deflection angle $  \alpha$ which verifies that the deflection angle reduces with the increase of ModMax nonlinear parameter. Regarding the influence of $F(R)$ gravity's parameter on the deflection angle, it can be seen from  Fig. \ref{Fig7}(c) that light rays are more deflected by BHs in $F(R)$ gravity compared to BHs in GR. Figs. \ref{Fig7}(d) and \ref{Fig7}(e) display the influence of curvature background on the deflection angle. As we see, the effect of $R_{0}$ parameter  on $\alpha$ depends on the value of the impact parameter $b$.  In dS spacetime, and for small (large) values of $b$, increasing $R_{0}$ leads to  increasing (decreasing) $\alpha$ (see Fig. \ref{Fig7}(d)), whereas the opposite behavior is observed in dS spacetime (see Fig. \ref{Fig7}(e)).  

\section{Conclusions}
\label{conclusion}
In this work, we presented an in-depth study of the optical features of BHs in $F(R)-$ModMax gravity, such as the shadow geometrical shape, energy emission rate, and
deflection angle. We first investigated the shadow observed by a distant observer and discussed how the shadow size is affected by the parameters of the model. According to our finding, the electric charge has a decreasing effect on the radius of the shadow, revealing the fact that the shadow size shrinks in the presence of the electric field. While both ModMax parameter and $F(R)$ gravity,s parameter have increasing contribution to the shadow size.

In the next step, we considered M87* BH as a model for  BHs in $F(R)-$ModMax theory and imposed some constraints on the modified gravity's parameters to have consistent results with observational data. Our findings show that the F(R) gravity parameter plays a crucial role in achieving results consistent with the EHT data. Studying the resulting shadow of dS (AdS) BHs in this modified gravity, we found that the shadow size is in agreement with the observational data of M87* for  $f_{R_{0}}>-1$ ($f_{R_{0}}<-1$).  Moreover, we computed the Schwarzschild shadow deviation ($\delta$) by using measurements of EHT and noticed that for large (small) values of the parameters of the model, a dS BH in $F(R)-$ModMax gravity with the same mass as Schwarzschild, has a greater (smaller) shadow size compared to the Schwarzschild BH. While an AdS BH has a greater shadow size compared to the Schwarzschild BH for large (small) values of the electric charge and ModMax parameter ($ \vert R_{0} \vert $ and $ \vert f_{R_{0}} \vert $).

Then, we continued by calculating the energy emission rate and examined the influence of
the BH parameters on the radiation process. Our results showed that the electric charge decreases the energy emission rate, indicating that the evaporation process becomes slower for a BH located in a more powerful electric field. Studying the effect of the ModMax parameter ($\gamma$), we noticed that the BH has a shorter lifetime in ModMax nonlinear electrodynamics theory. Similar to the ModMax parameter, $F(R)$ gravity's parameter has also an increasing effect on the emission rate, meaning that BHs in $F(R)$ gravity  have a shorter lifetime compared to BHs in GR. Regarding the influence of the curvature of the space-time on the evaporation process, we found that dS BHs have a longer lifetime compared to AdS BHs.

Finally, we performed an in-depth analysis of the gravitational lensing of light around such BHs. We noticed that an increase in electric charge leads to an increase in the deflection angle $\alpha$, while an increase in the ModMax parameter $\gamma$ causes a decrease in $\alpha$. This reveals the fact that light rays are deflected more (less) from the straight path in the presence of an electric field (ModMax field). We also studied the effect of the $F(R)$ gravity's parameter on the deflection angle and found that the light rays deviate more from their straight path in $F(R)$ gravity. In addition, our finding also showed that the influence of curvature background on the deflection angle depends on the value of the impact parameter $b$. In dS spacetime, increasing $R_{0}$ leads to  increasing (decreasing) $\alpha$ for small (large) values of $b$, whereas the opposite behavior will be observed in AdS spacetime.    

\begin{acknowledgements}
The authors would like to thank  Qiang Wu and Yiyang Wang for their fruitful comments and discussions which improved the
quality of the paper.

\end{acknowledgements}

\end{document}